\documentclass[twocolumn]{autart}

\usepackage{amsmath}
\usepackage{amssymb}
\usepackage{amsfonts}
\let\theoremstyle\relax
\usepackage{amsthm}

\usepackage{mathtools}
\MHInternalSyntaxOn
\MH_set_boolean_T:n {outer_mult}
\MHInternalSyntaxOff
\usepackage{tikz}
\usepackage{graphicx}
\usepackage{cancel}
\usepackage{textcomp}
\usepackage{adjustbox}
\usepackage{enumitem}
\usepackage{xcolor}
\usepackage{bbold}
\usepackage{newclude}
\usepackage{pgf}
\usepackage{tikz}
\usetikzlibrary{shapes,arrows,automata,graphs,positioning}
\usepackage{blindtext}
\usepackage{cite}

\newcommand{\diag}{\mathop{\mathrm{diag}}}
\renewcommand{\Re}{\mathrm{Re}}
\renewcommand{\Im}{\mathrm{Im}}

\renewcommand{\exp}{\mathrm{e}}

\newcommand{\tabref}[1]{Table~\ref{#1}}

\newcommand{\figref}[1]{Fig.~\ref{#1}}
\newcommand{\defref}[1]{Def.~\ref{#1}}
\newcommand{\corref}[1]{Cor.~\ref{#1}}
\newcommand{\lemref}[1]{Lemma~\ref{#1}}
\newcommand{\thmref}[1]{Theorem~\ref{#1}}
\newcommand{\secref}[1]{Section~\ref{#1}}
\newcommand{\assref}[1]{Assumption~\ref{#1}}
\newcommand{\exref}[1]{Example~\ref{#1}}

\newcommand{\shortref}[1]{\ref{#1}}
\newtheorem{theorem}{Theorem}[]
\newtheorem{definition}{Definition}[]
\newtheorem{assumption}{Assumption}[]
\newtheorem{corollary}{Corollary}[theorem]
\newtheorem{lemma}{Lemma}[theorem]
\newtheorem{remark}{Remark}[theorem]
\newtheorem{example}{Example}[theorem]
\begin{document}
	
\begin{frontmatter}

	\title{Frequency-Domain Modelling of Reset Control Systems using an Impulsive Description} 
	
	\thanks[footnoteinfo]{Corresponding author S. H. HosseinNia}
	
	\author[PME,DCSC]{R. N. Buitenhuis}\ead{R.N.Buitenhuis@student.tudelft.nl},    
	\author[PME]{N. Saikumar}\ead{N.Saikumar@tudelft.nl},               
	\author[PME]{S. H. HosseinNia}\ead{S.H.HosseinNiaKani@tudelft.nl} 
	
	\address[PME]{Precision and Microsystems Engineering, Faculty of Mechanical Engineering, Delft University of Technology, The Netherlands}                                       
	\address[DCSC]{Delft Center for Systems and Control, Faculty of Mechanical Engineering, Delft University of Technology, The Netherlands}             	
	
	
	\begin{abstract}
	\par The ever-increasing industry desire for improved performance makes linear controller design run into fundamental limitations. Nonlinear control methods such as Reset Control (RC) are needed to overcome these. RC is a promising candidate since, unlike other nonlinear methods, it easily integrates into the industry-preferred PID design framework. Thus far, RC has been analysed in the frequency domain either through describing function analysis or by direct closed-loop numerical computation. The former computes a simplified closed-loop RC response by assuming a sufficient low-pass behaviour. In doing so it ignores all harmonics, which literature has found to cause significant modelling prediction errors. The latter gives a precise solution, but by its direct closed-loop computation does not clearly show how open-loop RC design translates to closed-loop performance. The main contribution of this work is aimed at overcoming these limitations by considering an alternative approach for modelling RC using state-dependent impulse inputs. This permits accurately computing closed-loop RC behaviour starting from the underlying linear system, improving system understanding. A frequency-domain description for closed-loop RC is obtained, which is solved analytically by using several well-defined assumptions. This analytical solution is verified using a simulated high-precision stage, critically examining sources of modelling errors. The accuracy of the proposed method is further substantiated using controllers designed for various specifications.
\end{abstract}
\vspace{2mm}
	
\end{frontmatter}

\section{Introduction}
% Industry pushes limits
\par Industry is continuously pushing control limitations by increasing performance demands. This causes requirements on bandwidth, disturbance rejection, noise attenuation and reference tracking to become increasingly stringent. PID and other linear controllers are standard to industry, also to high-tech applications. This status is expected to prevail \cite{Samad2019}, because these controllers permit the industry preferred loop-shaping design framework. Linear control is inherently subject to fundamental limitations, including the Bode gain-phase relationship \cite{Bode1945}. This links bandwidth, disturbance rejection, noise attenuation and reference tracking. One cannot improve on some aspect without compromising on another. This design trade-off hinders the industry push for better performance.
% Nonlinear solutions, with RC unique!
\par This trade-off can only be overcome through nonlinear control, such as Reset Control (RC). RC is a promising candidate as various implementations embed nicely into PID and additionally the industry preferred loop-shaping framework. The first reset element was the Clegg Integrator (CI) \cite{Clegg1958}, which is an integrator with its state value resetting to zero whenever its input crosses zero. Through Describing Function (DF) analysis \cite{Gelb1968} it is shown that the CI inflicts $52^\circ$ less phase lag than the in gain similar linear integrator, thus overcoming the Bode gain-phase relationship.
% Expansion on RC
\par Several authors made contributions towards generalizing the CI. The first extension was by resetting a first-order low-pass filter known as the First Order Reset Element (FORE) \cite{Horowitz1975}. Further developments enhancing design flexibility include second-order \cite{Hazeleger2016} and fractional-order \cite{Saikumar2017} reset elements, as well as a second-order single-state reset element\cite{Karbasizadeh2020}. Additional tuning freedom was obtained by allowing states to be reset to non-zero values \cite{Beker2002,Guo2009a}. Recently, the Constant-in-Gain Lead-in-phase (CgLp) RC implementation was proposed \cite{Saikumar2019}, designed to provide a broadband phase lead without affecting the gain. This property makes CgLp very suitable to be used in combination with any linear controller.
% Reset Laws
\par The reset law accompanying a reset element determines when a reset occurs. Traditionally, that is when the input of the reset element crosses zero \cite{Clegg1958}. Extensions \cite{Vidal2008,Vidal2010} and alternatives \cite{Ne2005,Fichera2008,Ghaffari2014,Guo2009,Prieur2013,Li2013} are mentioned in literature, providing a variety of RC behaviours, tuning possibilities, stability results and performance analysis \cite{VanLoon2017,Journal2011}. These options are not considered here, as loop-shaping based RC tuning is developed for zero-crossing reset laws.
% Industry examples
\par Several works have demonstrated that RC can push performance beyond limits attainable through linear control \cite{Krishnan1974,Zheng2000,Beker2001a,Beker2001b}, for example by reducing overshoot \cite{Feuer1997} without affecting other specifications. RCs have been implemented in various control applications, including chemical processes \cite{Ghaffari2014}, vibration isolation \cite{Saikumar2019c} and motion control systems \cite{Banos2006,Guo2009,Palanikumar2018,Chen2018,Chen2019}.
% Analysis
\par A frequency-domain description of RC is imperative for design using the loop-shaping methodology preferred by industry. Most commonly, DF analysis is utilized \cite{Guo2009a}, which ignores all output harmonics. Despite this popularity, several works found DF to yield predictions in closed-loop that deviated widely from measurements \cite{Saikumar2019b,Saikumar2019c,Saikumar2019a}. Recently, an open-loop extension of DF analysis, incorporating harmonics termed higher-order sinusoidal input describing functions (HOSIDFs), was used together with various assumptions to compute a novel closed-loop frequency-domain description, CL-DF \cite{Saikumar2020}. DF assumes the reset element to have a sinusoidal input, while CL-DF assumes that the higher harmonics are small relative to the main harmonic, which at best holds approximately in closed-loop. Both methods also model two resets per input period only, known to not hold generally \cite{Banos2006}. Another frequency-domain method was suggested by \cite{Dastjerdi2020a}, which computes the closed-loop directly by solving numerically. This yields a precise solution at the cost of being computationally intensive and not providing a link between open- and closed-loop. None of the available methods sufficiently links open-loop RC design, especially considering the underlying linear system to closed-loop behaviour. Without such a link RC design is impaired, as it is not clear how certain tuning choices affect the closed-loop performance. This work aims to bridge this gap.
% Impulse idea
\par Some authors have mentioned that RC can be modelled as a linear controller with a train of state-dependent weighted impulse inputs \cite{Krishnan1974,Hollot2001,Chait2002}, but this idea is only developed for a CI \cite{Krishnan1974} and for certain nonlinear systems \cite{Haddad2000}, and not with the objective to find a frequency-domain solution. This work takes the impulsive RC modelling and generalizes that to obtain a closed-loop frequency domain description of RC systems, exploiting the resulting linear controller model, enabling accurate computation of closed-loop solutions in a way compatible to the industry preferred loop-shaping methodology. This accurately connects open-loop RC design to its closed-loop performance.
% Sections
\par The remainder of this paper is structured as follows. First, preliminaries of RC, including reset elements, definitions and stability results are given in \secref{sec:prelim}. Existing frequency domain analysis methods are presented and evaluated in \secref{sec:fdom}. \secref{sec:impulse} introduces the impulse formulation for a general RC in open-loop, irrespective of reset law, followed by a closed-loop formulation. This impulsive modelling is then formulated in the frequency domain in \secref{sec:per}, which is essential for industry. Only systems with zero-crossing reset laws are considered from there on. This modelling is simplified in \secref{sec:sol} using clearly stated assumptions to allow for an analytical solution. \secref{sec:setup} states the setup used to examine the effects of these assumptions in \secref{sec:val}. Afterwards, the accuracy is evaluated in \secref{sec:res} by using controllers tuned for various specifications. Last, \secref{sec:con} concludes this paper.

\section{Preliminaries on Reset Control}
\label{sec:prelim}
\par This section presents a generic reset control framework, along with related definitions and a stability theorem. 
\subsection{Reset control}
\begin{figure}[t]
	\centering
	\includegraphics[width=\linewidth]{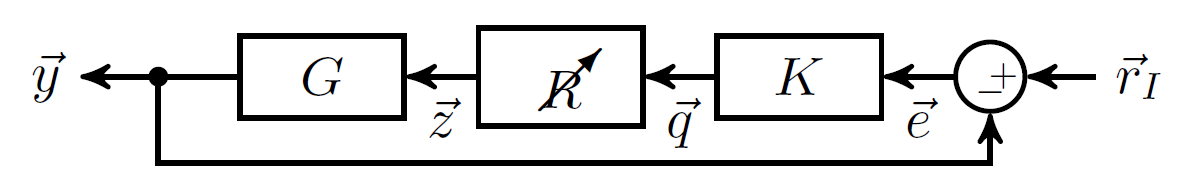}
	\caption[Closed-loop block diagram with a reset controller]{Block diagram of a reset control system with input $\vec{r}_I$ and output $\vec{y}$, consisting of reset element $\cancelto{}{R}$ surrounded by linear systems $K$ and $G$.}
	\label{blck:CL}
\end{figure}

Consider the generic setup given in \figref{blck:CL}, consisting of linear systems $K$ and $G$ surrounding reset element $\cancelto{}{R}$, with input $\vec{r}_I(t)$ and output $\vec{y}(t)$. Let $\vec{y}(t),\,\vec{e}(t),\,\vec{r}_I(t)\:{\in}\:\mathbb{R}^{m_y}$, $\vec{z}(t)\:{\in}\:\mathbb{R}^{m_z}$ and $\vec{q}(t)\:{\in}\:\mathbb{R}^{m_q}$, with ${m_y},\,{m_z},\,{m_q}\:{\in}\:\mathbb{N}$.
\begin{definition}[Reset Controller (RC)]\label{def:ODE}
	\par Let a reset controller $\cancelto{}{R}\:{:}\:\vec{q}(t)\:{\mapsto}\:\vec{z}(t)$ be defined by dimension-compatible matrices $A_R,\,B_R,\,C_R,\,D_R$, states $\vec{x}(t)\:{\in}\:\mathbb{R}^{n_{ol}},\,n_{ol}\:{\in}\:\mathbb{N}$ and reset matrix $A_{\rho}$. A reset occurs when $t\;{=}\;t_{r}\:{\in}\:t_{R}$, where $t_{R}$ is the set of all reset instants. The following equations describe $\cancelto{}{R}$:
	\begin{align}
	\label{eq:ODE}
	\cancelto{}{R}\;:\;\begin{cases}
	\dot{\vec{x}}(t)\;{=}\;A_R\,\vec{x}(t)\:{+}\:B_R\,\vec{q}(t),\ &t\:{\notin} \:t_R\\
	\vec{x}^+(t)\;{=}\;A_{\rho,r}\,\vec{x}(t),\ & t\:{\in}\:t_{R}\\
	\vec{z}(t)\;{=}\;C_R\,\vec{x}(t)\:{+}\:D_R\,\vec{q}(t) &
	\end{cases}
	\end{align}
	After-reset states are denoted by $\vec{x}^+$. Description \eqref{eq:ODE} permits a MIMO RC and any arbitrary reset law.
\end{definition}
\begin{definition}[Reset types] \label{def:BLS}
	\par In literature, reset matrix $A_{\rho,r}$ is generally diagonal and can also be non-constant. A fixed matrix is considered in this work: $A_{\rho}\;{=}\;\diag\left(\gamma_{1},\dots,\,\gamma_{n_{ol}}\right)$, with values $\gamma_{i}\:\in\:\left[{-}\:1,\,1\right]$, $i\:{\in}\:\left\{1,\dots,\,n_{ol}\right\}$. Define the following:
	\begin{itemize}[noitemsep]
		\item Fully resetting RC: $\gamma_{i}\;{\in}\;\left\{0,\,1\right\}{,}\ {\forall}\:i$ and ${\exists}\:i\:{\vert}\:\gamma_i\;{=}\;0$.
		\item Partially resetting RC: $A_{\rho}$ where ${\exists}\:i\:{\vert}\:\gamma_{i}\;{\notin}\;\left\{0,\,1\right\}$.
	\end{itemize}
\end{definition}
\begin{definition}[Reset control system (RCS)] \label{def:CL}
	\par Let the closed-loop reset control system $\cancelto{}{T}\:{:}\:\vec{r}_I(t)\:{\mapsto}\:\vec{y}(t)$ as in \figref{blck:CL} be defined by dimension-compatible matrices $A_{cl},\,B_{cl},\,C_{cl},\,D_{cl}$ and $A_{\rho cl}$, with states $\vec{x}_{cl}\:{\in}\:\mathbb{R}^{n_{cl}}$, $n_{cl}\:{\in}\:\mathbb{N}$. $\cancelto{}{T}$ is described by:
	\begin{align}
	\label{eq:CL}
	\cancelto{}{T}\;:\;\begin{cases}
	\dot{\vec{x}}_{cl}(t)\;{=}\;A_{cl}\,\vec{x}_{cl}(t)\:{+}\:B_{cl}\,\vec{r}_I(t),\ &t\:{\notin} \:t_R\\
	\vec{x}_{cl}^+(t)\;{=}\;A_{\rho cl}\,\vec{x}_{cl}(t),\ & t\:{\in}\:t_{R}\\
	\vec{y}(t)\;{=}\;C_{cl}\,\vec{x}_{cl}(t)\:{+}\:D_{cl}\,\vec{r}_I(t)&
	\end{cases}
	\end{align}	
\end{definition}
\par RCs as in \eqref{eq:ODE}, \eqref{eq:CL} are SISO if $m_y\;{=}\;1$, $m_z\;{=}\;1$ and $m_y\;{=}\;1$.
\par Note that henceforth RC refers to the open-loop nonlinear controller and RCS refers to the closed-lop reset system.
\begin{definition}[Base-Linear System (BLS)] \label{def:BLS2}
	\par The base-linear system of \figref{blck:CL} is obtained by removing all resets from the RCS, rendering it linear. The BLS sensitivity function $S_L(s)$ and complementary sensitivity function $T_L(s)$ are given by:
	\begin{align}
	S_L(s)\;&{\triangleq}\;\left(I\:{+}\:G(s)\,R_L(s)\,K(s)\right)^{-1}
	\label{eq:S}\\
	T_L(s)\;&{\triangleq}\;G(s)\,R_L(s)\,K(s)\,\left(I\:{+}\:G(s)\,R_L(s)\,K(s)\right)^{-1}
	\end{align}
	where $R_L$ denotes $\cancelto{}{R}$ without reset action:
	\begin{align}
	R_L(s)\;{\triangleq}\;C_R\,(sI\:{-}\:A_R)^{-1}\,B_R\:{+}\:D_R
	\end{align}
\end{definition}
\begin{definition}[Zero-crossing law] \label{def:0inreset}
	\par A SISO RC with zero crossing law resets when $t_R\;{=}\;\left\{t\:{\in}\:\mathbb{R}\:{\vert}\:\vec{q}(t)\;{=}\;0\right\}$.
\end{definition}
\begin{definition}[Time regularization] \label{def:timereg}
	\par Time regularization suppresses any reset if $t\;<\:t_p\:+\:\tau$, with $\tau\:{>}\:0$ a tunable parameter and $t_p$ the last occurred reset time instant \cite{Ne2005}.
\end{definition}
\par RC systems can be prone to deadlock, beating and Zeno behaviour \cite{Banos2006}, causing solutions to be ill-defined. This behaviour can be avoided by using time regularization \cite{Zaccarian2005,Ne2005}. Any discrete-time implementation inherently features time regularization, having $\tau$ equal to the sampling time \cite{Heertjes2016}. As most practical implementations are discretized it is chosen to disregard deadlock, beating and Zenoness in this paper. It is assumed that solutions to \eqref{eq:CL} are well-defined.

\subsection{Reset elements}
\par
\begin{figure}[t]
	\centering
	\includegraphics[width=\linewidth]{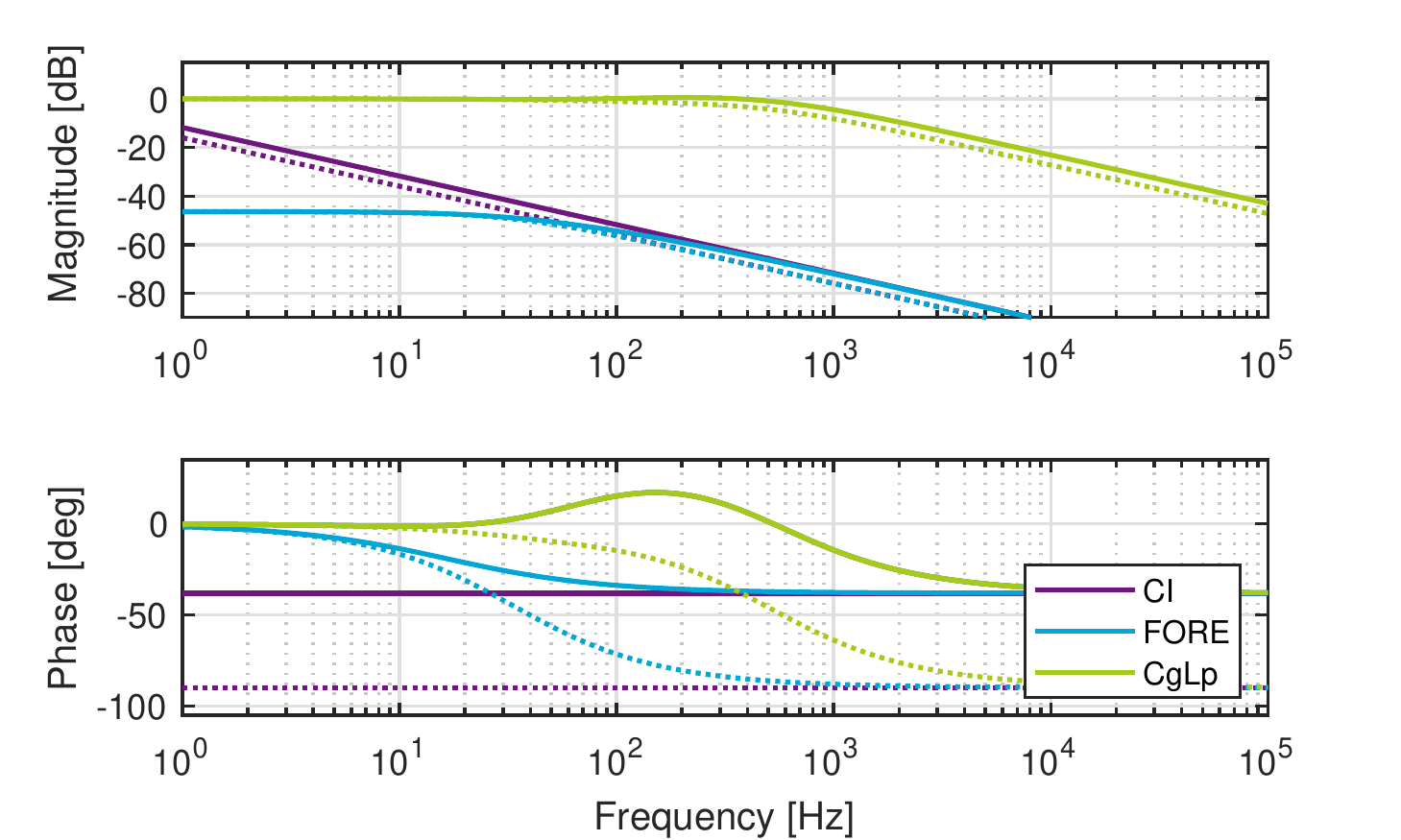}
	\caption{Bode Plot with the linear, $R_L$ (dashed line), and first harmonic, $R_{DF}$ (solid line), responses for a CI, FORE and CgLp reset element. The harmonic responses, computed through DF analysis \eqref{eq:HOSIDF1}, use full reset ($\gamma\;{=}\;0$).}
	\label{fig:bodereset} 
\end{figure}

 Various reset elements are presented in literature. A few relevant ones are given below, all of them most commonly using zero-crossing reset laws. \figref{fig:bodereset} gives the Bode plots for these, depicting their base-linear and nonlinear first harmonic responses.
\vspace{1mm}
\subsubsection{Generalized Clegg Integrator (GCI)}
\par The Clegg Integrator \cite{Clegg1958} is a resetting integrator, which can be generalized by allowing partial resets through $\gamma$. The GCI is defined with:
\begin{align}
A_R\;{=}\;0,\ B_R\;{=}\;1\,\ C_R\;{=}\;1,\ D_R\;{=}\;0,\ A_\rho\;{=}\;\gamma \label{eq:CI}
\end{align}
\subsubsection{Generalized First Order Reset Element (GFORE)}
\par FORE is based on a first order low-pass filter. It was first given by \cite{Horowitz1975} and later generalized by permitting partial resets \cite{Guo2009}. A GFORE with corner frequency $\omega_{r}$ is given with:
\begin{align}
A_R\;{=}\;-\omega_{r},\ B_R\;{=}\;\omega_{r},\ C_R\;{=}\;1,\ D_R\;{=}\;0,\ A_\rho\;{=}\;\gamma \label{eq:FORE}
\end{align}
\subsubsection{Constant in gain, Lead in phase (CgLp)}
\par CgLp is a novel RC element providing broadband phase lead while maintaining unit gain \cite{Saikumar2019b}. This characteristic enables CgLp to be combined with any linear controller, increasing phase without inflicting gain alterations. This is achieved by merging a GFORE, having pole $\omega_{r\alpha}$, with a lead-lag filter, having pole $\omega_f$ and zero $\omega_{r}\;{=}\;\omega_{r\alpha}\,\alpha$. Parameter $\alpha$ corrects for the GFORE pole shift induced by reset nonlinearity \cite{Saikumar2019b}, ensuring that the GFORE pole remains coincident with the lead-lag zero.
\begin{align}
\left[
\begin{array}{c|c}
A_R & B_R \\ \hline
C_R & D_R
\end{array}\right]\;{=}\;&\left[\begin{array}{cc|c} 
{-}\:\omega_{r\alpha} & 0 & \omega_{r\alpha}   \label{eq:CgLp} \\
\omega_f &  {-}\:\omega_f &  0    \\ \hline
\omega_f\:{/}\:\omega_r & 1\:{-}\:\omega_f\:{/}\:\omega_r & 0
\end{array}\right]\\
\quad A_\rho\;{=}\;&\diag\left[\gamma,\,1\right] \nonumber
\end{align}
\subsection{Stability}
\par Consider a SISO RCS where the matrices $A_{cl}$, $C_{cl}$, $A_{\rho cl}$ and $A_{\rho}$ can be structured as below. This factorization is always possible if $G(s)$ has no direct feed-through.
%\par Hbeta evaluates stability for \figref{blck:hbeta}, whilst we use \figref{blck:CL}. Autonomous $y$ \figref{blck:CL} is stable if $q$ \figref{blck:hbeta} is stable, given that $K$ does not cancel RHP poles of $G$ or $R$. Use combined state-space matrices of $A_{KG}$, $B_{KG}$, $C_{KG}$ and $D_{KG}$.
\begin{align}
\begin{split}
A_{cl}\;&{=}\;\left[\begin{array}{cc} \bullet & \bullet \\ \bullet & A_R \end{array} \right]\\
C_{cl}\;&{=}\;\left[\begin{array}{cc} C_G & 0 \end{array} \right]
\end{split}\quad\begin{split}
A_{\rho cl}\;&{=}\;\left[\begin{array}{cc} I_{n_{cl}-n_{ol}} & 0 \\ 0 & A_{\rho} \end{array} \right]\\
A_{\rho}\;&{=}\;\left[\begin{array}{cc} I_{\bar{\rho}} & 0 \\ 0 & A_{\rho}^\star \end{array} \right]
\end{split}\nonumber
\end{align}
where $\bullet$ denotes the appropriate matrix. $A_{\rho}^\star\:{\in}\:\mathbb{R}^{n_{\rho}\times n_\rho}$, $n_\rho\:{\in}\:\mathbb{N}_0$ is a matrix corresponding to the  $n_\rho$ resetting states. It follows that the number of non-reset states is $n_{\bar{\rho}}\;{=}\;n_{ol}\:{-}\:n_\rho$. $C_G$ is the corresponding $C$ matrix of linear system $G$ %Define the number of added closed-loop states $n_c\;{=}\;n_{cl}\:{-}\:n_{ol}$.
\begin{theorem}[$\mathcal{H}_\beta$ - condition]
	\par An autonomous SISO RCS \eqref{eq:CL} with zero-crossing reset law is said to satisfy the $\mathcal{H}_\beta$ condition if ${\exists}\:\beta\:{\in}\:\mathbb{R}^{n_\rho}$, $P_{\rho}\:{\in}\:\mathbb{R}^{{n_\rho}\,{\times}\,{n_\rho}}\;{>}\;0$ such that
	\begin{align}
	\mathcal{H}_\beta\;{\triangleq}\;\begin{bmatrix}
	\beta\,C_{P}&0_{n_\rho\:{\times}\:n_{\bar{\rho}}}&P_{\rho}
	\end{bmatrix}\,\left(sI\:{-}\:A_{cl}\right)^{-1}\,\begin{bmatrix}
	0\\0_{n_{\bar{\rho}}\:{\times}\:n_\rho}\\I_{n_\rho\:{\times}\:n_\rho}
	\end{bmatrix} \nonumber
	\end{align}
	is strictly positive real, $A_{\rho cl}$ is non-zero and \cite{Guo2016}:
	\begin{align}
	{A_{\rho}^\star}^TP_{\rho}\,A_{\rho}^\star\:{-}\:P_{\rho}\;{\leq}\;0 \nonumber
	\end{align}
	\par The SISO RCS \eqref{eq:CL} with a zero-crossing reset law is quadratically stable if and only if it satisfies $\mathcal{H}_\beta$ condition \cite{Beker2002}. Uniformly exponential convergence and input-to-state convergence also hold if \eqref{eq:CL} satisfies $\mathcal{H}_\beta$ \cite{Dastjerdi2020a}.
\end{theorem}
\section{Frequency-domain describing methods}
\label{sec:fdom}
\par The available frequency-domain methods for describing RCs are given next. These methods are applicable to zero-crossing resets only. \tabref{tab:assumptions1} provides an overview of the various assumptions used by methods describing RCSs.

\subsection{DF and HOSIDF analysis}
\par Analysing SISO RCs in the frequency domain is typically performed using DF analysis, which computes the first harmonic in the Fourier series expansion of $\vec{z}(t)$. This requires \eqref{eq:ODE} to have a globally asymptotically stable $2\,\pi\,{/}\,\omega$ - periodic output $\vec{z}(t)$, for a sinusoidal input $\vec{q}(t)$ with frequency $\omega\;{>}\;0$. This happens if and only if \cite{Guo2009a}:
\begin{align}
\left\vert\lambda\left(A_\rho\,e^{A_R\,\delta}\right)\right\vert\;{<}\;1,\quad {\forall}\:\delta\:{\in}\:\mathbb{R}^+  \label{eq:OLreq} 
\end{align}
\par DF analysis is extended to Higher Order Sinusoidal Input Describing Function (HOSIDF) analysis by also considering the harmonics in the Fourier series expansion of $\vec{z}(t)$.  \cite{Saikumar2020}.
\begin{theorem}[DF \cite{Guo2009a}, HOSIDF \cite{Heinen2018}] The $n$-th order HOSIDF for an open-loop SISO RC \eqref{eq:ODE} satisfying \eqref{eq:OLreq} with zero-crossing resets and a sinusoidal input with frequency $\omega\;{>}\;0$ is computed by:
\begin{align}
&R_{DF,n}(\omega)\;{\triangleq}\;C_R(j\omega n\,I\:{-}\:A_R)^{-1}\nonumber\\
&\qquad {\times}\:\begin{cases}
\left(I\:{+}\:j\theta_D(\omega)\right)\,B_R\:{+}\:D_R,\quad &n\:{=}\:1\\
j\theta_D(\omega)\,B_R,&\text{odd }n\:{>}\:1\\
0,&\text{even }n\:{>}\:1
\end{cases} \label{eq:HOSIDF1}
\end{align}
\vspace{-4mm}
\begin{align*}
\text{where:}\quad\!\theta_D(\omega)\;&{\triangleq}\;{-}\:\frac{2\omega^2}{\pi}\,\Delta(\omega)\left[\Gamma_R(\omega)\:{-}\:\Lambda^{-1}(\omega)\right]\qquad\\
\Gamma_R(\omega)\;&{\triangleq}\;\Delta_R^{-1}(\omega)\,A_\rho\,\Delta(\omega)\,\Lambda^{-1}(\omega)\\
\Lambda(\omega)\;&{\triangleq}\;\omega^2\,I\:{+}\:A_R^2\\
\Delta(\omega)\;&{\triangleq}\;I\:{+}\:\text{e}^{\frac{\pi}{\omega}\,A_R}\\
\Delta_R(\omega)\;&{\triangleq}\;I\:{+}\:A_\rho\,\text{e}^{\frac{\pi}{\omega}\,A_R}
\end{align*}
\par  DF analysis equals \eqref{eq:HOSIDF1} for $n\;{=}\;1$. The corresponding DF-approximated sensitivity function for \eqref{eq:CL} is:
\begin{align}
S_{DF}(\omega)\;&{=}\;\left(I\:{+}\:G(\omega)\,R_{DF,1}(\omega)\,K(\omega)\right)^{-1} \label{eq:Sdf}
\end{align}
\par This approximation assumes (i) that all harmonics are negligible in closed-loop, (ii) $\vec{q}(t)$ to be sinusoidal, which (iii) implicitly assumes the RCS to have two resets per input period.
\end{theorem}
\par
\begin{figure}[t]
	\centering
	\includegraphics[width=\linewidth]{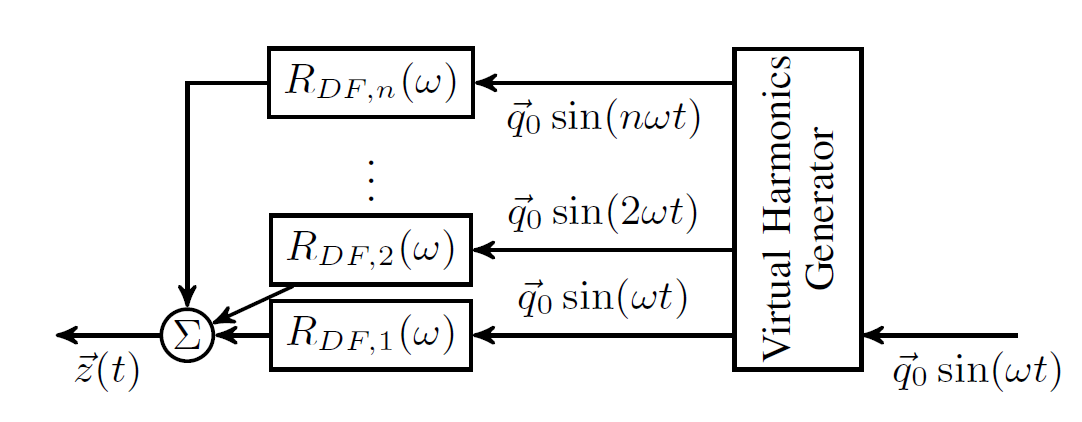}
	\caption[HOSIDF block diagram]{HOSIDF representation of $\cancelto{}{R}$ for a sinusoidal input $\vec{q}(t)$, using a virtual harmonics generator. Adapted from \cite{Nuij2006}.}
	\label{blck:HOSIDF}
\end{figure}

 HOSIDF analysis models the RC response in open-loop for input $\vec{q}(t)\;=\;\vec{q}_0\,\sin(\omega t)$, as shown in \figref{blck:HOSIDF}.
\begin{align}
Z(\omega)\;&{=}\;\sum\nolimits_{n=1}^\infty R_{DF,n}(\omega)\,Q(\omega)\;{\Leftrightarrow}\;\cancelto{}{R}(s)\,Q(s) \label{eq:zhosidf}\\
\vec{z}(t)\;&{=}\;\sum\nolimits_{n=1}^\infty\left\vert R_{DF,n}\,\vec{q}_0\right\vert \,\mathrm{e}^{j\angle R_{DF,n}\,{+}\,jn\omega t}
\end{align}
\par Assumptions for $S_{DF}$ do not hold. Reset induces harmonics, which through feedback prevent $\vec{q}(t)$ from being fully sinusoidal. RCSs often have more than two resets per period \cite{Banos2006}.
\begin{table}[t]
	\centering
	\caption{Overview of assumptions existing methods for computing frequency-domain closed-loop RC behaviour use. Empty fields indicate that there are no assumptions.}
	\label{tab:assumptions1}
	\begin{tabular}{lccc}
		\hline 
		& & & \\[-0.8em]
		& DF   & CL-DF      & CL-FR        \\ \hline
		& & & \\[-0.8em]
		Modelled resets per period:     & $2$           & $2$         &          \\
		Signals assumed sinusoidal:           & $\vec{q}(t)$  & $\vec{r}_I(t)$& $\vec{r}_I(t)$\\
		Resets assumed at:         &             & $\vec{q}_{DF,1}\;{=}\;0$  &  \\
		Neglects harmonics:         & Yes            &            &           \\ \hline
	\end{tabular}
\end{table}
\subsection{Closed-loop HOSIDF analysis}
\par Recently a method was presented that extends HOSIDF to RCS \cite{Saikumar2020}. Starting from open-loop HOSIDF, this method assumes (i) that there are exactly two resets per input period, spaced $\pi\,/\,\omega_r$ apart, and (ii) that solely the first harmonic, $Q(\omega)\,{=}\,K(\omega)\,S_{DF,1}(\omega)$, causes and affects resets.
\begin{theorem}[Closed-loop HOSIDF (CL-DF) \cite{Saikumar2020}]
	\par The $n$-th order CL-DF for an RCS with an input-to-state convergent SISO RC \eqref{eq:ODE} satisfying \eqref{eq:OLreq}, having zero-crossing resets and a sinusoidal input with frequency $\omega\:{>}\:0$, is defined by:
	\begin{align}
	S_{DF_{CL},n}(\omega)\,{\triangleq}\,
	\begin{cases}
	Sl_1(\omega),\,&n\,{=}\,1\\
	{-}\,Sl_{bls}(n\omega)\,L_n(\omega)\,Sl_{1,n}(\omega),\,& n\,{>}\,1
	\end{cases}
	\end{align}
	\vspace{-4mm}
	\begin{align}
	\text{Where: }\qquad L_n(\omega)\;&{\triangleq}\;G(n\omega)\,R_{DF,n}(\omega)\,K(\omega) \; \qquad \qquad \nonumber\\
	Sl_n(\omega)\;&{\triangleq}\;\left(I\:{+}\:L_n(\omega)\right)^{-1} \nonumber\\
	L_{bls}(\omega)\;&{\triangleq}\;G(\omega)\,R_L(\omega)\,K(\omega) \nonumber\\
	Sl_{bls}(\omega)\;&{\triangleq}\;\left(I\:{+}\:L_{bls}(\omega)\right)^{-1} \nonumber\\
	Sl_{1,n}(\omega)\;&{\triangleq}\;\left(\left\vert Sl_1(\omega)\right\vert\exp\left(jn\angle Sl_1(\omega)\right)\right) \nonumber
	\end{align}
\end{theorem}
\par CL-DF uses assumptions to close the loop, and hence introduces errors in modelling and prediction, yet improves upon $S_{DF}$ as it includes harmonics. Additionally, by considering all harmonics including first together, it does not provide any link between the base-linear system and the introduction of reset.
\subsection{Closed-Loop Frequency Response (CL-FR)}
\par CL-FR is different from the other approaches mentioned, as it analyses a stable SISO RCS with zero crossing resets and a sinusoidal input directly through numerical evaluation \cite{Dastjerdi2020a}. This direct closed-loop computation yields accurate results at the cost of not providing insight in how open-loop RC design translates to RCS performance.
\section{Impulse Reset Modelling}
\label{sec:impulse}
\par RCs are modelled as linear systems with a state-dependent timed impulse train input, a description first mentioned by \cite{Krishnan1974}. The following theorem will prove that any general open-loop RC as in \eqref{eq:ODE} is equivalent to a linear system with impulse inputs. This result is then used to acquire the RCS description.

\begin{theorem}[Impulse-based RC modelling] \label{thm:impulsive} The states $\dot{\vec{x}}(t)$ of any open-loop RC as in \eqref{eq:ODE} are computed as:
	\begin{equation}
	\dot{\vec{x}}(t)\;{=}\;A_R\,\vec{x}(t)\:{+}\:B_R\,\vec{q}(t)\:{+}\:\sum_{t_{r}{\in}{t_{R}}}(A_{\rho}\:{-}\:I)\,\vec{x}(t)\,\delta(t_{r}) \label{eq:impulsive}
	\end{equation}
	\begin{proof}
		\par Consider an open-loop RC \eqref{eq:ODE}. The after-reset states $\vec{x}^+(t)$ at $t\;{=}\;t_{r}\:{\in}\:t_{R}$ are given by:
		\begin{align}
		\vec{x}^+(t_{r})\;&{=}\;A_{\rho}\,\vec{x}(t_{r})\;{\equiv}\;I\,\vec{x}(t_{r})\:{+}\:(A_{\rho}\:{-}\:I)\,\vec{x}(t_{r}) \nonumber
		\end{align}		
		\par The term $(A_{\rho}\,{-}\,I)\,\vec{x}(t_r)$ is added at a reset, $t\;{=}\;t_{r}$. This can be modelled as a Heaviside step function $H(t_{r})$. Doing so, differentiating and substituting $\dot{\vec{x}}(t)$ from \eqref{eq:ODE} gives:
		\begin{align}
		\vec{x}^+(t_{r})\;&{=}\;\vec{x}(t_{r})\:{+}\:(A_{\rho}\:{-}\:I)\,\vec{x}(t_{r})\,H(t_{r}) \nonumber\\
		\dot{\vec{x}}^+(t_{r})\;&{=}\;\dot{\vec{x}}(t_{r})\:{+}\:(A_{\rho}\:{-}\:I)\,\vec{x}(t_{r})\,\delta(t_r)\nonumber\\
		&{=}\;A_R\vec{x}(t_{r})\:{+}\:B_R\,\vec{q}(t_{r})\:{+}\:(A_{\rho}\:{-}\:I)\,\vec{x}(t)\,\delta(t_{r})\nonumber
		\end{align}
		where $\delta(t)$ is the Dirac delta function. It is noted that the reset only changes $\dot{\vec{x}}(t)$ at $t_{r}$. At $t\:\notin\:t_R$, the states thus flow according to $R_L$. Summing over $t_{r}\,\in\,t_{R}$ yields the result.
	\end{proof}
\end{theorem}

\begin{corollary}[RC Laplace formulation] \label{cor:laplace}
	\par The output and states of \eqref{eq:ODE} are given in the Laplace domain by:
	\begin{align}
	Z(s)\;&{=}\;R_L(s)\,Q(s)\:{+}\:R_{\delta}(s)\sum_{{t_{r}{\in}{t_{R}}}}\vec{x}(t_{r})\,\exp^{{-}\:t_{r}\,s} \label{eq:Z}\\
	X(s)\;&{=}\;R^X_{L}(s)\,Q(s)\:{+}\:R^X_{\delta}(s)\sum_{{t_{r}{\in}{t_{R}}}}\vec{x}(t_{r})\,\exp^{{-}\:t_{r}\,s} \label{eq:Xol}
	\end{align}
	where $R_{\delta}(s)$ is given by $R_{\delta}(s)\;{\triangleq}\;C_R\,(sI\:{-}\:A_R)^{-1}\,(A_{\rho}\:{-}\:I)$, transfer functions to $X(s)$ by $R^X_{L}(s)\;{\triangleq}\;(sI\:{-}\:A_R)^{-1}\,B_R$ and $R^X_{\delta}(s)\;{\triangleq}\;(sI\:{-}\:A_R)^{-1}\,(A_{\rho}\:{-}\:I)$.
	\begin{proof}
		\par Start by writing \eqref{eq:impulsive} in the Laplace domain:
		\begin{multline}
		sX(s)\:{=}\:A_R\,X(s)\:{+}\:B_R\,Q(s)\,{+}\sum_{{t_{r}{\in}{t_{R}}}}\!\left(A_{\rho,k}\,{-}\,I\right)\,\vec{x}(t_{r})\,\exp^{{-}\,t_{r}\,s} \nonumber
		\end{multline}
		Vector $\vec{x}(t_{r})$ is evaluated at a specific time instant and can therefore be treated as a constant. Rewriting for $X(s)$ and substitution of $R^X_{L}(s)$ and $R^X_{\delta,k}(s)$ gives \eqref{eq:Xol}. Solving for $X(s)$, using \eqref{eq:ODE} to write $Z(s)\;{=}\;C_R\,X(s)\:{+}\:D_R\,Q(s)$ and afterwards inserting \eqref{eq:Xol}, $R_{L}(s)$ and $R_{\delta}(s)$ yields \eqref{eq:Z}.
	\end{proof}			
\end{corollary}
\begin{figure}[b]
	\centering
	\includegraphics[width=\linewidth]{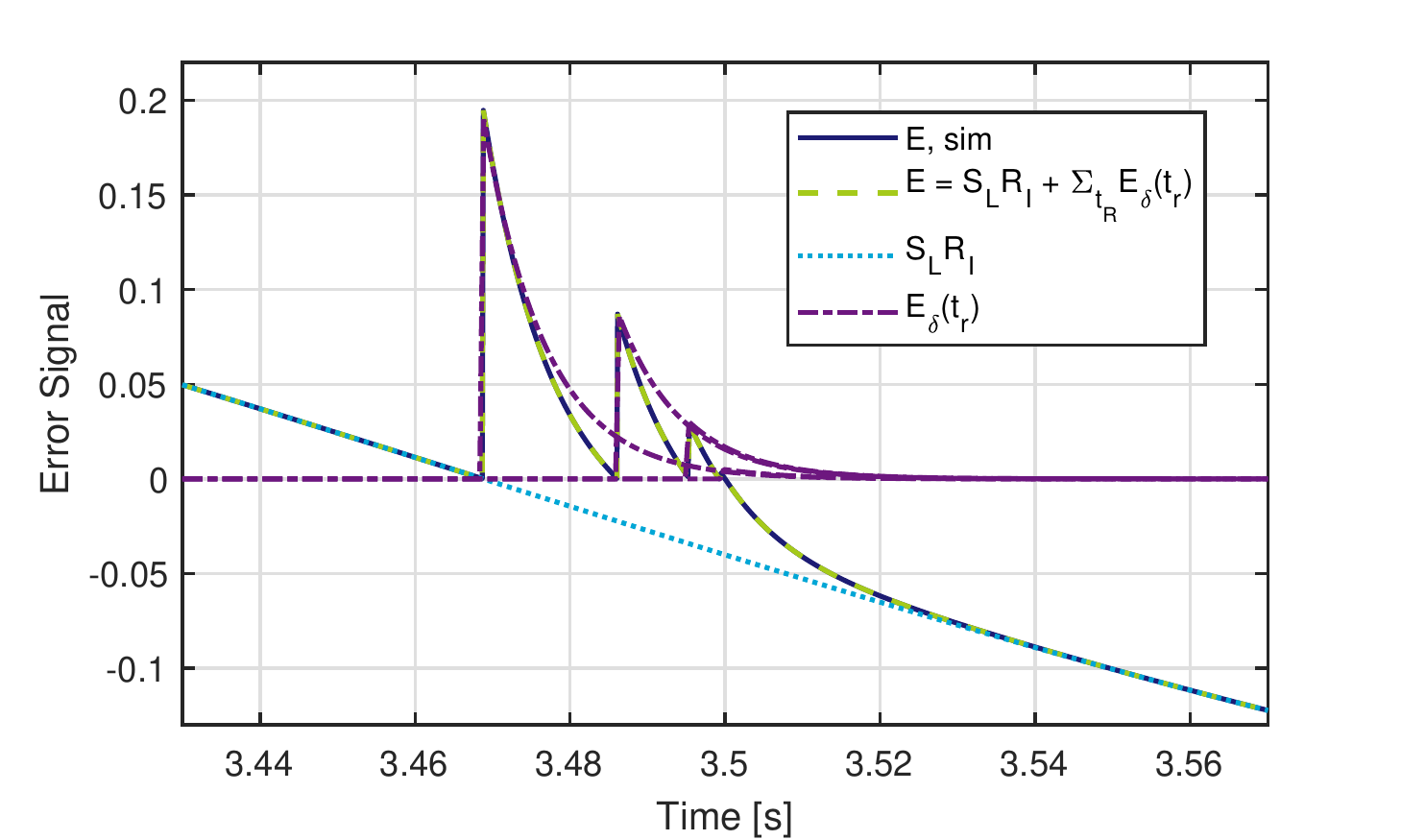}
	\caption{Detail on a series of resets around a zero crossing for a FORE closed-loop reset system as given by \exref{ex:impulse}, using $\vec{r_I}\;=\;\sin\left(t\:/\:2\pi\right)$. The components of \eqref{eq:E}, their sum and the simulated response are shown.}
	\label{fig:impulse} 
\end{figure}
\begin{corollary}[Closed-loop $E(s)$] \label{cor:cl}
	\par The RCS error response $E(s)$ is computed to be the BLS summed by impulse responses:
	\begin{multline}
	E(s)\;{=}\;S_L(s)\,R_I(s)\\ \:{-}\:S_L(s)\,G(s)R_{\delta}(s)\sum_{t_{r}{\in}{t_{R}}}\vec{x}(t_{r})\,\exp^{{-}\:t_{r}\, s}\label{eq:E}
	\end{multline}
	\par The second term is denoted as $E_\delta(s,t_r)$ to simplify notation, giving $E(s)\;{=}\;S_L(s)\,R_I(s)\:{+}\:\sum\nolimits_{t_{r}{\in}{t_{R}}} E_{\delta}(s,t_{r})$.
	\begin{proof}
		\par From \figref{blck:CL} it follows that $Q(s)\;{=}\;K(s)\,E(s)$. Together with \eqref{eq:Z} this gives:
		\begin{align}
		Z(s)\;&{=}\;R_L(s)\,K(s)\,E(s)\;{+}\;R_{\delta}(s)\sum_{t_{r}{\in}{t_{R}}}\vec{x}(t_{r})\,\exp^{{-}\:t_{r}\,s} \nonumber
		\end{align}
		\par In \figref{blck:CL} it is seen that $E(s)\;{=}\;R_I(s)\:{-}\:G(s)\,Z(s)$:
		\begin{multline}
		E(s)\;{=}\;R_I(s)\:{-}\:G(s)\,R_L(s)\,K(s)\,E(s)\\
		\:{-}\:G(s)R_{\delta}(s)\sum_{t_{r}{\in}{t_{R}}}\vec{x}(t_{r})\,\exp^{{-}\:t_{r}\,s} \nonumber
		\end{multline}
		\par The result follows by solving for $E(s)$ and inserting \eqref{eq:S}.
	\end{proof}
\end{corollary}

\begin{corollary}[Closed-Loop $X(s)$] \label{cor:x}
	\par The closed-loop states of $\cancelto{}{R}(s)$ are computed to be:
	\begin{multline}
	X(s)\;{=}\;R^X_{L}(s)\,K(s)\,S_L(s)\,R_I(s)\\
	\:{-}\:R^X_{L}(s)\,K(s)\,S_L(s)\,G(s) R_{\delta}(s)\sum_{t_{r}{\in}{t_{R}}}\vec{x}(t_{r})\,\exp^{{-}\:t_{r}\,s}\\
	\:{+}\:R^X_{\delta}(s)\sum_{t_{r}{\in}{t_{R}}}\vec{x}(t_{r})\,\exp^{{-}\:t_{r}\,s} \label{eq:X}
	\end{multline}
	\begin{proof}
		Take \eqref{eq:Xol} and substitute $Q(s)$ with $K(s)E(s)$, using the RCS error $E(s)$ as defined by \eqref{eq:E}.
	\end{proof}
\end{corollary}
\begin{remark} \label{rem:xcl}
	\par All results based on \thmref{thm:impulsive} require uniqueness and existence of a solution to \eqref{eq:ODE} only. No requirements on input types, system dimensions, stability or reset types are needed to compute the response. The RCS behaviour can thus be computed exactly, given that reset times $t_{R}$ and corresponding states $\vec{x}(t_r)$ are known.
\end{remark}
\begin{remark} \label{rem:mimo}
	\par These results accept MIMO systems. However, MIMO reset control implementations often use multiple reset conditions \cite{Paesa2011,Zhao2018}. The obtained results permit a straightforward extension to an arbitrary number of reset conditions, where each of these corresponds to some reset matrix and resets at some subset of reset times $t_R$. It follows that these results can describe closed-loop MIMO RC behaviour.
\end{remark}
\begin{example} \label{ex:impulse}
	\par Consider a SISO FORE in closed-loop with a zero-crossing reset law, using $K(s)\;{=}\;100$, $G(s)\;{=}\;1$, $\omega_r\;{=}\;25$ and $\gamma\;{=}\;0$. \figref{fig:impulse} illustrates in time-domain how the linear response $S_L(s)\,R_I(s)$ and impulse responses $E_{\delta}(s,t_{r})$ are summed to create \eqref{eq:E}, which equals the simulated response.
\end{example}
\begin{remark} \label{rem:insight}
	\par Result \eqref{eq:E} adds insight into RCS performance by linking how the base-linear system designed in open-loop and the introduction of reset in the form of impulses affects closed-loop performance. In closed-loop the RC behaves as the BLS, but having impulse responses with tunable weighting $I\:{-}\:A_\rho$ added to it. Thus, the closed-loop can be estimated by considering the BLS and weighted impulse response based on the open-loop design. This analysis allows to explain in a different way why RCs are found to have a lower sensitivity peak than their corresponding BLSs \cite{Saikumar2019}. From \eqref{eq:E} it follows that this must occur because impulse responses partially cancel out the BLS error.
\end{remark}
\section{Periodic results}
\label{sec:per}
\par A precise RCS solution is obtained if the values $t_{R}$ and $\vec{x}(t_r)$ are computed. It might be possible to obtain these in a numerical manner to yield a precise and generic solution, without needing further assumptions or setup requirements. However, such a solution does not generally permit a frequency-domain description, which is imperative for loop-shaping. Periodicity is used to rewrite \eqref{eq:E} in frequency-domain terms for zero-crossing reset systems.
\subsection{Periodicity of RC}
\begin{figure}[t]
	\centering
	\includegraphics[width=\linewidth]{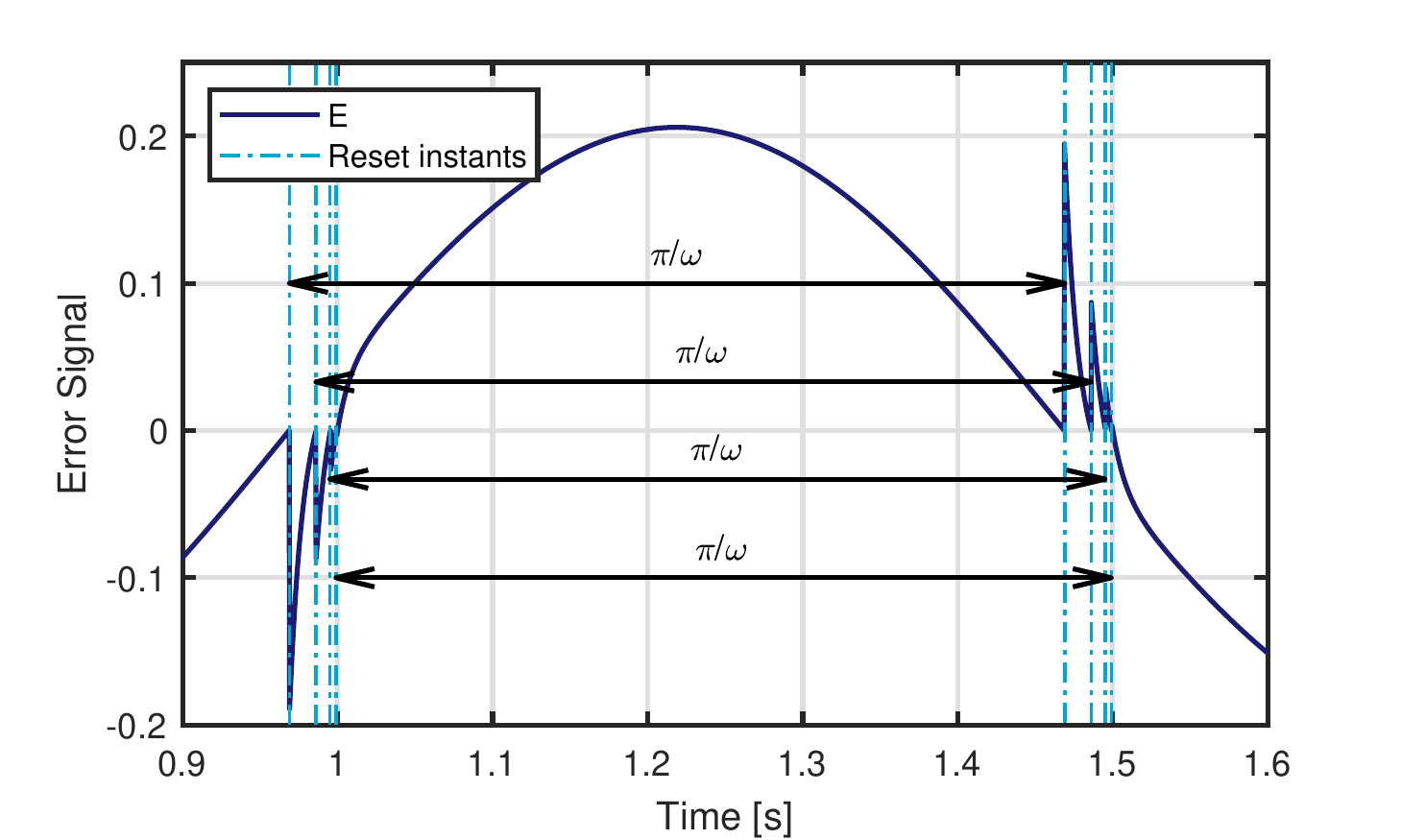}
	\caption{Time response of a FORE RCS system satisfying \thmref{thm:per} with indicated time intervals between resets, for $\vec{r}_I\;{=}\;\sin\left(t\:/\:2\pi\right)$.}
	\label{fig:periodic}
\end{figure}

\begin{theorem}[Periodic RC \cite{Dastjerdi2020a}]\label{thm:per}
	\par If a SISO RCS \eqref{eq:CL} with zero-crossing law (a) satisfies $\mathcal{H}_\beta$, and (b) has a purely sinusoidal $\vec{r}_I(t)$ with frequency $\omega$, then, in steady-state, the RCS has (i) a unique periodic solution $\vec{x}_{cl}(t)$, $\vec{y}(t)$ with period $2\,\pi\:{/}\:\omega$, (ii) all even harmonics equal to zero, and (iii) a periodic pattern of reset instants with period $\pi\:{/}\:\omega$.
\end{theorem}
\begin{remark}
	\par \cite{Dastjerdi2020a} \thmref{thm:per} also holds if, instead of \eqref{eq:CL} satisfying $\mathcal{H}_\beta$, it is Uniformly Bounded Steady-State.
\end{remark}
\begin{example} \label{ex:per}
	\par \figref{fig:periodic} gives the steady-state time response of the RCS given by \exref{ex:impulse} to a $1$ Hz sinusoidal reference. This setup meets the requirements of \thmref{thm:per}, which therefore predicts that reset instants have a $\pi\,{/}\,\omega$ periodic pattern. \figref{fig:periodic} illustrates that this holds.
\end{example}
\begin{corollary}[Periodic impulse response] \label{cor:h}
	\par Define a new set of reset times, $t\;{=}\;t_{r}\:{\in}\:t_{\rho}$ with $t_{\rho}\;{=}\;\left\{t\:{\in}\:t_{R}\:{\vert}\:t\:{\in}\:[0,\,\pi\:{/}\:\omega\rangle \right\}$. If \thmref{thm:per} is satisfied the following simplification holds:
	\begin{align}
	&R_{\delta}(s)\sum_{t_{r}\in t_R}\vec{x}(t_{r})\,\exp^{{-}\:t_{r}\,s}\;{=}\;\sum_{t_r\in t_{\rho}}\xi\left(s,t_{r},\vec{x}(t_{r})\right) \label{eq:h}
	\end{align}
	The term $\xi\left(s,t_{r},\vec{x}(t_{r})\right)$ is by definition $2\,\pi\:{/}\:\omega_r$ periodic:
	\begin{multline}
	\xi\left(s,t_{r},\vec{x}(t_{r})\right)\;{\triangleq}\;R_{\delta}(s)\!\sum_{p\:{\in}\:2\mathbb{Z}}\!\left(\vec{x}(t_{r})\,\exp^{{-}\:(t_{r}{+}p\frac{\pi}{\omega})\,s}\right.\\
	\left.{-}\:\vec{x}(t_{r})\,\exp^{{-}\:(t_{r}{+}(p{+}1)\,\frac{\pi}{\omega})\,s}\right) \nonumber
	\end{multline}
	\begin{proof}
		\par If \thmref{thm:per} is satisfied, the reset instants are $\pi\,{/}\,\omega_r$ periodic. As such, $t_\rho$ can be used to represent all resets, $\cup_{p}\{t_\rho\:{+}\:p\,\pi\:{/}\:\omega_r\}\;{=}\;t_R,\ p\;{\in}\;\mathbb{Z}$. These are thus equal:
		\begin{align}
		\sum_{t_r\in t_{R}}\vec{x}(t_{r})\,\exp^{{-}\:t_{r}\,s}\;{=}\;\!\sum_{t_r\in t_{\rho}}\sum_{p \in\mathbb{Z}}\vec{x}(t_{r}\:{+}\:p\,\pi\:{/}\:\omega)\,\exp^{{-}\:(t_{r}{+}p\frac{\pi}{\omega})\,s} \nonumber
		\end{align}
		\par Using \thmref{thm:per}, $\vec{x}(t_{r}\:{+}\:p\,\pi\:{/}\:\omega)$ can be expressed in $\vec{x}(t_{r})$.
		\begin{multline}
		\sum_{t_r\in t_{R}}\vec{x}(t_{r})\,\exp^{{-}\:t_{r}\,s}\;{=}\;{\sum_{t_r\in t_{\rho}}}{\sum_{p\:{\in}\:2\mathbb{Z}}}\!\left(\vec{x}(t_{r})\,\exp^{{-}\:(t_{r}+p\frac{\pi}{\omega})\,s}\right.\\
		\left.{-}\:\vec{x}(t_{r})\,\exp^{{-}\:(t_{r}+(p+1)\frac{\pi}{\omega})\,s}\right) \nonumber
		\end{multline}
		\par Pre-multiplication with $R_{\delta}(s)$ and inserting $\xi\left(s,t_{r},\vec{x}(t_{r})\right)$ as defined above completes the proof.
	\end{proof}
\end{corollary}
\subsection{Impulse HOSIDF analysis}
\par HOSIDF describes the open-loop response, as given by \eqref{eq:impulsive}, thus modelling both the periodic impulse responses and the linear response to the input. For convenience in further derivations a new HOSIDF definition is proposed which solely models the impulse responses. This new definition requires a sinusoidal input, as conventional for HOSIDF.
\begin{definition}[Impulse HOSIDF]\label{def:hosidf2}
	\par The $n$-th order impulse HOSIDF analysis for a SISO RC \eqref{eq:ODE} satisfying \eqref{eq:OLreq} with zero-crossing resets, given a sinusoidal input with frequency $\omega\;{>}\;0$, is defined as a function of input matrix $B^\star$:
	\begin{align}
	&R_{DF,n}^\star(\omega,B^\star)\;{\triangleq}\;C_R\,(j\omega\,n\,I\:{-}\:A_R)^{-1}\, \nonumber\\
	&\qquad{\times}\:\begin{cases}
	\,j\theta_D(\omega)B^\star,\quad &\text{odd }n\:{>}\:0\\
	0,&\text{even }n\:{>}\:1
	\end{cases} \label{eq:HOSIDF2}
	\end{align}
	\par If $B^\star\;{=}\;B_R,$ the two HOSIDF formulations obey:
	\begin{align}
	\sum_{n=1}^\infty R_{DF,n}(\omega)\;{=}\;R_L(\omega)\:{+}\:\sum_{n=1}^\infty R^\star_{DF,n}(\omega,\,B_R) \label{eq:rihosidf}
	\end{align}
\end{definition}
\subsection{Closed-loop frequency-domain description}
\par This section shows that open-loop Impulse HOSIDF can be used to exactly model RCSs. A virtual input $Q^\star$ and input matrix $B^\star$ are computed, which cause the open-loop Impulse HOSIDF response to exactly model that of the RCS \eqref{eq:CL}. The open-loop states are first established for Impulse HOSIDF.
\begin{lemma}
	\par The following is used to simplify results:
	\begin{align}
	\Re\left\{(j\omega\,I\:{-}\:A_R)^{-1}j\right\}\;&{\equiv}\;\left(\omega^2\,I\:{+}\:A_R^2\right)^{-1}\omega\,I \label{eq:RLXj}\\
	\Re\left\{(j\omega\,I\:{-}\:A_R)^{-1}\right\}\;&{\equiv}\;{-}\:\left(\omega^2\,I^2\:{+}\:A_R^2\right)^{-1}A_R \label{eq:RLX}
	\end{align}
	%\begin{proof}
%		\par Take the left-hand sides of \eqref{eq:RLXj}, multiply by $({-}\:j\omega\,I\:{-}\:A_R)^{-1}\,({-}\:j\omega\,I\:{-}\:A_R)\;{=}\;I$, and simplify:
%		\begin{align}
%		&{=}\;\Re\left\{(j\omega\,I\:{-}\:A_R)^{-1}({-}\:j\omega\,I\:{-}\:A_R)^{-1}({-}\:j\omega\,I\:{-}\:A_R)\,j\right\} \nonumber\\
%		&{=}\;\Re\left\{(\omega^2\,I\:{+}\:A_R^2)^{-1}(\omega\,I\:{-}\:j\,A_R)\right\} \nonumber
%		\end{align}
%		\par Taking the same approach for \eqref{eq:RLX} yields:
%		\begin{align}
%		&{=}\;\Re\left\{(j\omega\,I\:{-}\:A_R)^{-1}({-}\:j\omega\,I\:{-}\:A_R)^{-1}({-}\:j\omega\,I\:{-}\:A_R)\,\right\} \nonumber\\
%		&{=}\;\Re\left\{(\omega^2\,I\:{+}\:A_R^2)^{-1}({-}\:j\omega\,I\:{-}\:A_R)\right\} \nonumber
%		\end{align}
%		\par Extracting real parts gives the desired results.
%	\end{proof}
\end{lemma}
\begin{theorem}[Open-loop states $\vec{x}({t_{r}})$] \label{lem:xhosidf}
	\par Consider an open-loop SISO RC \eqref{eq:ODE} using zero-crossing resets with a sinusoidal input $\vec{q}(t)$ having amplitude $\vec{q}_0\:{\in}\:\mathbb{R}$ and frequency $\omega$. Define reset sets based on the derivative $\dot{\vec{q}}(t)$:
	\begin{align*}
	t_{r}^\downarrow\:{\in}\:t_{R}^\downarrow\;{=}\;\{t\:{\in}\:t_{R}\: {:} \:\dot{\vec{q}}(t)\:{<}\: 0\}\\
	t_{r}^\uparrow\:{\in}\:t_{R}^\uparrow\;{=}\;\{t\:{\in}\:t_{R}\: {:} \:\dot{\vec{q}}(t)\:{>}\:0\}
	\end{align*}  
	so that $t^\downarrow_{R}\:{\cup}\:t^\uparrow_{R}\;{=}\;t_{R}$. The states $\vec{x}(t^\downarrow_{r})$ obey:
	\begin{multline}
	\vec{x}(t^\downarrow_{r})\;{=}\;\left(I\:{-}\:\left(I\:{+}\:\exp^{A_R\frac{\pi}{\omega}}A_\rho\right)^{-1}\,\exp^{A_R\frac{\pi}{\omega}}\,\left(A_\rho\:{-}\:I\right)\right)\\
	{\times}\:\left(\omega^2\,I\:{+}\:A_R^2\right)^{-1}\,\omega\,I\,B_R\,\vec{q_0} \label{eq:xhosidf}
	\end{multline}
	where $(I\:+\:\exp^{A\,\frac{\pi}{\omega}}A_\rho)$ is assumed to be invertible.
	\begin{proof}
		\par Split the open-loop states \eqref{eq:Xol} in two parts, so that $X(s)\;{=}\;X_L(s)\;{+}\;X_\delta(s)$:
		\begin{align}
		X_L(s)\;&{=}\;R^X_{L}(s)\,Q(s)\nonumber \\
		X_\delta(s)\;&{=}\;R^{X}_{\delta}(s)\sum_{t_r\in t_{R}}\vec{x}(t_{r})\,\exp^{{-}\:t_{r}\,s} \nonumber
		\end{align}		
		\par First consider the linear term, $X_L(s)$. States $\vec{x}_L(t)$ can be obtained by taking the real part of $X_L(s)\;{=}\;X_L(j\omega)$ evaluated at some time instance. This is possible because of linearity combined with having a sinusoidal input. Rewriting the sinusoidal input $\vec{q}(t)$ gives:
		\begin{align}
		\vec{q}(t)\;&{=}\;\vec{q}_0\,\sin(\omega t)\;{=}\;\Im\{\vec{q}_0\,\exp^{j\omega t}\}\;{=}\;\Re\{\vec{q}_0\,\exp^{j\omega (t-\frac{\pi}{2})}\}\nonumber 
		\end{align}
		\par This is used to write $X_L(j\omega)$ in time domain:
		\begin{align}
		X_L(j\omega)\;&{=}\;(j\omega I\:{-}\:A_R)^{-1}B_R\,Q(j\omega)\nonumber\\
		\vec{x}_L(t)\;&{=}\;\Re\{(j\omega I\:{-}\:A_R)^{-1}B_R\}\vec{q}(t)\; \nonumber\\
		&{=}\;\Re\{(j\omega I\:{-}\:A_R)^{-1}B_R\,\vec{q}_0\,\exp^{j\omega (t-\frac{\pi}{2})}\}\nonumber
		\end{align}
		\par The zero-crossing reset law is used to determine $\vec{x}_L(t^\downarrow_{r})$, which requires finding $\vec{q}(t^\downarrow_r)$. A zero-crossing of $\vec{q}(t)$ implies $\Re\{\vec{q}(t_r)\}\;{=}\;0\;{\Leftrightarrow}\;\vec{q}_0\,\exp^{j\omega (t-\frac{\pi}{2})}\;{=}\;\pm j\vec{q}_0$. As $\omega\,{>}\,0$, function $\vec{q}_0\,\exp^{j\omega (t-\frac{\pi}{2})}$ propagates counter-clockwise, implying that solution ${+}\:j\vec{q}_0$ occurs when sinusoid $\vec{q}(t)$ crosses $0$ from above ($\dot{\vec{q}}(t)\:{\leq}\: 0$). Applying this and using \eqref{eq:RLXj} to simplify gives:
		\begin{align}
		\vec{x}_L(t^\downarrow_{r})\;&{=}\;\Re\left\{(j\omega\,I\:{-}\:A_R)^{-1}B_R\,j\,\vec{q}_0\right\}\nonumber\\
		\vec{x}_L(t^\downarrow_{r})\;&{=}\;\left(\omega^2\,I\:{+}\:A_R^2\right)^{-1}\,\omega\,I\,B_R\,\vec{q}_0\nonumber
		\end{align}
		\par The impulsive part is considered next. Write in time domain:
		\begin{align}
		\vec{x}_\delta(t)\;&{=}\;\!\sum_{t_r\in t_{{R}_{\leq t}}}\!\!\exp^{A_R(t-t_r)}(A_\rho\:{-}\:I)\,\vec{x}(t_r) \nonumber
		\end{align}
		\par From $\vec{q}(t)$ it follows that resets are spaced $\pi\,{/}\,\omega$ apart. Thus,  ${\forall}\:t^\downarrow_{r}\:{\in}\;t_R^\downarrow,\:{\exists}\:\tilde{t}_r^\uparrow\:{\in}\:t_{R}^\uparrow\: {\vert} \: t_{r}^\downarrow\:{-}\:\tilde{t}_{r}^\uparrow\;{=}\;\pi\:{/}\:\omega$. Evaluating at $t_r^\downarrow$ and expressing in terms of the states at the previous reset, $x_\delta(\tilde{t}_r^\uparrow)$:
		\begin{multline}
		\vec{x}(t_r^\downarrow)\;{=}\;\exp^{A_R(t_r^\downarrow-\tilde{t}_r^\uparrow)}(A_\rho\:{-}\:I)\,\vec{x}(\tilde{t}_r^\uparrow)\:\\{+}\:\exp^{A_R(t_r^\downarrow-\tilde{t}_r^\uparrow)}\!\sum_{t_r\in t_{{R}_{\leq \tilde{t}_r^\uparrow}}}\!\!\exp^{A_R(\tilde{t}_r^\uparrow-t_r)}(A_\rho\:{-}\:I)\,\vec{x}(t_r) \nonumber
		\end{multline}
		\vspace{-2mm}
		\par The last term is per definition equal to $x_\delta(\tilde{t}_r^\uparrow)$. Therefore,
		\begin{align}
		\vec{x}(t_r^\downarrow)\;&{=}\;\exp^{A_R(t_r^\downarrow-\tilde{t}_r^\uparrow)}(A_\rho\:{-}\:I)\,\vec{x}(\tilde{t}_r^\uparrow)\:{+}\:\exp^{A_R(t_r^\downarrow-\tilde{t}_r^\uparrow)}\,\vec{x}_\delta(\tilde{t}_r^\uparrow) \nonumber\\
		\vec{x}(t_r^\downarrow)\;&{=}\;\exp^{A_R\frac{\pi}{\omega}}\left((A_\rho\:{-}\:I)\,\vec{x}(\tilde{t}_r^\uparrow)\:{+}\:\vec{x}_\delta(\tilde{t}_r^\uparrow)\right) \nonumber		
		\end{align}
		\par If \eqref{eq:OLreq} is satisfied, as required for HOSIDF analysis, $\vec{x}(t)\,{=}\,{-}\:\vec{x}(t+\pi\,/\,\omega)$ \cite{Guo2009a}. Inserting $t_r^\downarrow$, $\tilde{t}_r^\uparrow$ gives $\vec{x}(t_r^\downarrow)\,{=}\,{-}\:\vec{x}(\tilde{t}_r^\uparrow)$. Expanding $\vec{x}(t)$ shows that\\ $\vec{x}(t_r^\downarrow)\;{=}\;\vec{x}_L(t_r^\downarrow)\:{+}\:\vec{x}_\delta(t_r^\downarrow)\:{=}\:-\vec{x}_L(\tilde{t}_r^\uparrow)\:{-}\:\vec{x}_\delta(\tilde{t}_r^\uparrow)$. From $\vec{x}_L(t_r)$ having $\pm\,j\vec{q}_0$ solutions on alternating reset times $\vec{x}_L(\tilde{t}_r^\downarrow)\;{=}\;{-}\:\vec{x}_L(t_r^\uparrow)$ follows. Thus, $\vec{x}_\delta(t_r^\downarrow)\;{=}\;{-}\:\vec{x}_\delta(t_r^\uparrow)$. Inserting this and writing for $\vec{x}(t_r^\downarrow)$ gives:
		\begin{align}
		\vec{x}_L(t_r^\downarrow)\:{+}\:\vec{x}_\delta(t_r^\downarrow)\;&{=}\;{-}\:\exp^{A_R\frac{\pi}{\omega}}\left((A_\rho\:{-}\:I)\,\vec{x}_L(t_r^\downarrow)\:{+}\:A_\rho\,\vec{x}_\delta(t_r^\downarrow)\right) \nonumber		
		\end{align}
		\par Solving for $\vec{x}_\delta(t_r^\downarrow)$ in terms of $\vec{x}_L(t_r^\downarrow)$ and inserting that in $\vec{x}(t_r^\downarrow)\;{=}\;\vec{x}_L(t_r^\downarrow)\:{+}\:\vec{x}_\delta(t_r^\downarrow)$ yields the desired solution.
	\end{proof}
\end{theorem}
\begin{remark} \label{rem:test}
	\par These states $\vec{x}_\delta(t_r^\downarrow)$ \eqref{eq:xhosidf} equal those of the Impulse HOSIDF case, given that $B_R\;{=}\;B^\star$ and $Q(\omega)\;{=}\;Q^\star(\omega)$.
\end{remark}
\par
\begin{figure}[t]
	\centering
	\includegraphics[width=\linewidth]{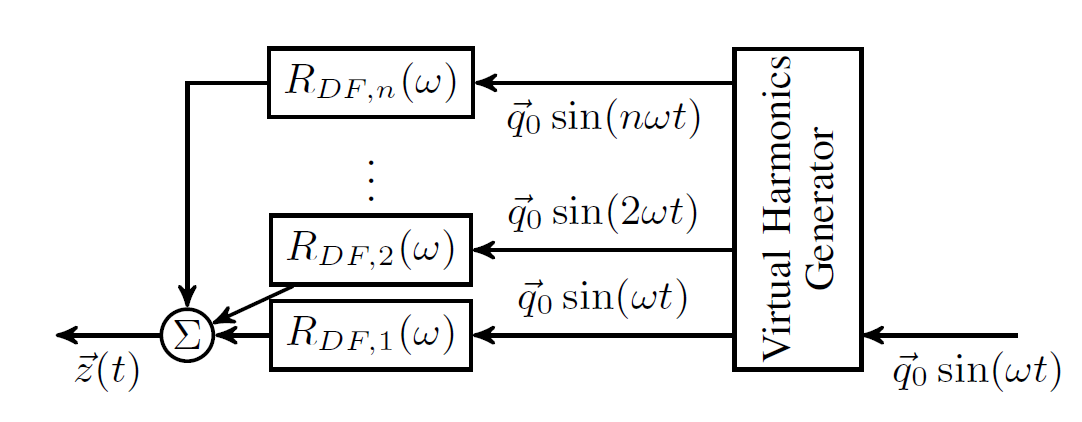}
	\caption[Block diagram representing \eqref{eq:Ehosidf}]{Block diagram representing \eqref{eq:Ehosidf}. The dashed area must be summed over all $t_r\:{\in}\:t_\rho$. Arguments of $R^\star_{DF,n}(\omega,B^\star(\omega,\vec{x}(t^\downarrow_{r})))$ are dropped.}
	\label{blck:Ehosidf}
\end{figure}

 Next, the virtual input $Q^\star(\omega)$ to the Impulse HOSIDF and corresponding input matrix $B^\star$ are computed as a function of $\vec{x}(t^\downarrow_{r})$. These are then used to find the closed-loop formulation.
\begin{corollary}[HOSIDF can model any $\vec{x}(t^\downarrow_{r})$] \label{lem:bstar}
	\par Impulse HOSIDF \eqref{eq:HOSIDF2} can model any periodic impulse response with states $\vec{x}(t^\downarrow_{r})$ by choosing the virtual input magnitude $\vec{q}^\star_0(\omega,\vec{x}(t^\downarrow_{r}))\;{=}\;1$ and the input matrix $B^\star(\omega,\vec{x}(t^\downarrow_{r}))$ as:
	\begin{align}
	B^\star(\omega,\vec{x}(t^\downarrow_{r}))\;&{=}\;\zeta(\omega)\,\vec{x}(t^\downarrow_{r}) \label{eq:qdf}
	\end{align}
	Where $\zeta(\omega)$ is defined as:
	\begin{align}
	\zeta(\omega)\;&{=}\;\begin{multlined}[t]\left(\left(\omega^2\,I\:{+}\:A_R^2\right)^{-1}\omega\,I\right)^{-1}\\
	{\times}\:\left(I\:{-}\:\left(I\:{+}\:\exp^{A_R\frac{\pi}{\omega}}A_\rho\right)^{-1}\!\exp^{A_R\frac{\pi}{\omega}}\left(A_\rho\:{-}\:I\right)\right)^{-1}\end{multlined} \nonumber
	\end{align}
	\begin{proof}
		\par Substitution of $B^\star(\omega,\vec{x}(t^\downarrow_{r}))$ for $B_R$ in \eqref{eq:xhosidf} while using virtual input $Q^\star(\omega,\vec{x}(t^\downarrow_{r}))$ with magnitude $\vec{q}^\star_0(\omega,\vec{x}(t^\downarrow_{r}))$ instead of $Q(\omega)$, and rewriting for $B^\star(\omega,\vec{x}(t^\downarrow_{r}))\,\vec{q}^\star_0(\omega,\vec{x}(t^\downarrow_{r}))$ afterwards, yields the result. This shows that by computing $B^\star(\omega,\vec{x}(t^\downarrow_{r}))$ any periodic reset state can be created, thus that Impulse HOSIDF can model any periodic impulse response.
		\par $B^\star(\omega,\vec{x}(t^\downarrow_{r}))$ makes magnitude $\vec{q}^\star_0(\omega,\vec{x}(t^\downarrow_{r}))$ obsolete, which is why $q^\star_0(\omega,\vec{x}(t^\downarrow_{r}))\;{=}\;1$ is chosen. The phase of $Q^\star(\omega)$ determines reset times, which is covered in a later section.
	\end{proof}
\end{corollary}
\begin{remark}
	\par Solutions to \eqref{eq:qdf} require $\omega\:{>}\:0$ as well as $\left(I\,{+}\,\exp^{A_R\frac{\pi}{\omega}}A_\rho\right)^{-1}\!\!\exp^{A_R\frac{\pi}{\omega}}\left(A_\rho\,{-}\,I\right)\,{\neq}\,I$, which hold generally.
\end{remark}
\begin{theorem}[HOSIDF analysis in closed-loop] \label{thm:hosidf}
	\par A summation of Impulse HOSIDF responses on top of the BLS can accurately describe $e(t)$ for any system satisfying \thmref{thm:per}, given continuity of $\vec{e}(t)$ ($\vec{e}(t)\:{\in}\:C^0$):
	\begin{multline}
	E(\omega)\;{=}\;S_L(\omega)\,R_I(\omega)\:{-}\:\sum_{n=1}^\infty\!S_L(n\omega)\,G(n\omega)\\ {\times}\:\sum_{t_{\rho}}\! R^\star_{DF,n}(\omega,B^\star(\omega,\vec{x}(t^\downarrow_{r})))\,Q^\star(\omega,\vec{x}(t^\downarrow_{r}))  \label{eq:Ehosidf}
	\end{multline}
	\par With $Q^\star(\omega,\vec{x}(t^\downarrow_{r}))$ having magnitude $1$ and phases to ensure the correct reset times. \figref{blck:Ehosidf} represents \eqref{eq:Ehosidf} graphically.
	\begin{proof}
		\par Take \eqref{eq:zhosidf} and insert \eqref{eq:rihosidf}, with a designable input matrix $B_R\;{=}\;B^\star(\omega,\vec{x}(t^\downarrow_{r}))$ and sinusoidal input $Q^\star(\omega,\vec{x}(t^\downarrow_{r}))$ with amplitude $1$. Arguments of $B^\star$ and $Q^\star$ are dropped.
		\begin{align}
		Z(\omega)\;{=}\;R_L(\omega)\,Q^\star(\omega)\:{+}\:\sum_{n=1}^\infty R^\star_{DF,n}(\omega,B^\star)\,Q^\star \nonumber
		\end{align}
		\par Inserting the open-loop $Z(\omega)$ \eqref{eq:Z}, using $B^\star$ and $Q^\star$ computed according to the actual states $\vec{x}(t^\downarrow_{r})$, and rewriting gives:
		\begin{align}
		\sum_{n=1}^\infty\! R^\star_{DF,n}(\omega,B^\star)\,Q^\star\;{=}\;
		R_{\delta}(\omega)\!\sum_{t_r\in t_{R}}\!\vec{x}(t_{r})\,\exp^{{-}\:t_{r}j\omega} \nonumber
		\end{align}
		\par The reset times $t\;{=}\;t_{r}\:{\in}\:t_{R}\;{=}\;\cup_{p}\{t_{r}\:{+}\:p\,\pi\:{/}\:\omega\},\ p\;{=}\;\mathbb{Z}$ follow from zero-input resets. Using these yields:
		\begin{multline}
		\sum_{n=1}^\infty\! R^\star_{DF,n}(\omega,B^\star)\,Q^\star\;{=}\; R_{\delta}(\omega)\,\sum_{p\:{\in}\:\mathbb{Z}}\vec{x}(t_{r}\:{+}\:p\,\pi\:{/}\:\omega)\\\,{\times}\:\exp^{{-}\:\left(t_{r}\:{+}\:p\,\pi\:{/}\:\omega\right)j\omega} \nonumber
		\end{multline}
		\par The correct HOSIDF response for resets at $t_\rho$ with weight $\vec{x}(t_r)$ is ensured if $B^\star$ and $Q^\star$ are computed according to \thmref{lem:bstar}. HOSIDF has odd harmonics only, thus $\vec{x}(t_r)\;{=}\;{-}\:\vec{x}(t_r\:{+}\:\pi\:{/}\:\omega)$. Simplifying using periodicity and inserting the definition of $\xi\left(s,t_{r},\vec{x}(t_{r})\right)$ from \corref{cor:h} gives:
		\begin{align*}
		\sum_{n=1}^\infty\!&\begin{multlined}[t]R^\star_{DF,n}(\omega,B^\star)\,Q^\star\;{=} R_{\delta}(\omega)\sum_{p\:{\in}\:2\mathbb{Z}}\left(\vec{x}(t_{r})\,\exp^{{-}\:(t_{r}{+}p\frac{\pi}{\omega})\,j\omega}\:\right.\\
		 \left.{-}\:\vec{x}(t_{r})\,\exp^{{-}\:(t_{r}{+}(p{+}1)\,\frac{\pi}{\omega})\,j\omega}\right)\end{multlined}\\
		\sum_{n=1}^\infty\!&R^\star_{DF,n}(\omega,B^\star)\,Q^\star\;{=}\;\xi\left(\omega,t_{r},\vec{x}(t_{r})\right)
		\end{align*}
		\par Thus far, this holds for open-loop Impulse HOSIDF. Conveniently, $\xi\left(s,t_{r},\vec{x}(t_{r})\right)$ as in \eqref{eq:h} is obtained, which can be substituted in the closed-loop response \eqref{eq:E}, which requires inserting the closed-loop values for $t_r$ and $\vec{x}(t_r)$.
	\end{proof}
\end{theorem}
\begin{remark}
	\par \thmref{thm:hosidf} proves that any SISO RC \eqref{eq:CL} with (i) zero-crossing resets (ii) satisfying the $\mathcal{H}_\beta$ condition and (iii) having $\vec{e}(t)\:{\in}\:C^0$ can be described without error by a summation of Impulse HOSIDF responses, if the set $t_{\rho}$ and states $\vec{x}(t_r)$ are known ${\forall}\:t_r\:{\in}\:t_{\rho}$.
\end{remark}
\begin{remark}
	\par \thmref{thm:hosidf} also holds if $\vec{e}(t)\:{\notin}\:C^0$, except for prediction errors caused by the Gibbs phenomenon in the vicinity of discontinuities.
\end{remark}
\par With \eqref{eq:Ehosidf} an accurate description is provided that can model the same RCSs in frequency-domain as CL-FR can, given that $t_{\rho}$ and $\vec{x}(t_r)$ are known, without needing further conditions.

%\begin{remark}
%	\par Invertibility of $(I\:{+}\:\exp^{A_R\,\frac{\pi}{\omega}}\,A_\rho)$ holds for most RCs, with the exception of a CI when $\gamma\;{=}\;{-}\:1$.
%\end{remark}
\section{Analytical solution}
\label{sec:sol}
\par To solve \eqref{eq:Ehosidf} time instants $t_{r}$ and states $\vec{x}(t_r)$ must be computed ${\forall}\:t_r\:{\in}\:t_{\rho}$. This can be done numerically or analytically. For the sake of simplifying analysis and staying close to what is familiar to the linear control/based loop-shaping methodology it is chosen to pursue an analytical solution. This choice comes at the cost of requiring assumptions, impacting precision.
\begin{assumption}[Two resets per period] \label{ass:xtr}
	\par Assume that sufficient accuracy is achieved by modelling a closed-loop RC \eqref{eq:CL} satisfying \thmref{thm:hosidf} with exactly two resets per period, taking $\vert t_\rho \vert\;{=}\;1$. Accurate modelling of the two modelled resets requires that any unmodelled reset does not significantly influence their position nor states.
\end{assumption}

\begin{assumption}[Zero crossing direction] \label{ass:rup}
	\par It is assumed that at any reset the direction of $\vec{q}(t)$ crossing zero is equal to that predicted by the nearby BLS zero crossing: $ \mathrm{sign}(\dot{\vec{q}}(t))\;{=}\;\mathrm{sign}(\dot{\vec{q}}_L(t)),\ {\forall}\:t\:{\vert}\:\vec{q}(t)\;{=}\;0$.	This holds for most RCSs and can be verified analytically.
\end{assumption}

\begin{assumption}[Existence of a reset instant]\label{ass:3}
	\par Assume that, at any reset instant, the absolute combined magnitude of all prior resets-induced impulse responses is less than the absolute peak value of the BLS response. This must hold in any real system satisfying \thmref{thm:per}. Otherwise, there cannot be a reset at $\pi\:{/}\:\omega$ distance from the previous reset, which contradicts \thmref{thm:per}. However, this assumption may be violated in cases where other assumptions affect solutions.
\end{assumption}

\begin{lemma}[Convergence] \label{lem:convergence}
	\par The series $\sum_{p\,{\in}\,\mathbb{N}}\exp^{Ap}$, with $A$ square, is convergent if all $\lambda\,(A)\:{<}\:0$.
	\begin{proof}
		\par Start by rewriting the problem into a Neumann series:
		\begin{align}
		\sum_{p\,{\in}\,\mathbb{N}}\exp^{A\,p}\;{=}\;\sum_{p\,{\in}\,\mathbb{N}}\left(\exp^{A}\right)^p \nonumber
		\end{align}
		\par This series converges if the spectral radius satisfies $\rho\,(\exp^A)\:{<}\:1$, thus if the eigenvalues $\lambda^\star$ corresponding to $\exp^A$ satisfy $\max\left\vert\lambda^\star\right\vert\;{<}\;1$. Because $\lambda^\star\;{=}\;\exp^\lambda$, where $\lambda$ denotes the eigenvalues of $A$, series convergence follows if ${\nexists}\;\lambda(A)\:{\geq}\:0$.	
		%Expressing $\lambda^\star$ in terms of the eigenvalues of $A$, denoted by $\lambda$, gives: 
		 %\begin{align}
		 %A\,v\;&{=}\;\lambda\,v\;{\rightarrow}\;A^2\,v\;{=}\;A\,(\lambda\,v)\;{=}\lambda^2\,v\;{\rightarrow}\;A^3\,v\;{=}\;\lambda^3\,v\nonumber\\
		 %\exp^A\,v\;&{=}\;\lambda^\star\,v\;{=}\;I\,v\:{+}\:A\,v\:{+}\:\tfrac{1}{2}\,A^2\,v\:{+}\:\tfrac{1}{6}\,A^3\,v\:{+}\:\dots\nonumber\\
		 %\lambda^\star\,v\;&{=}\;\left(1\:{+}\:\lambda\:{+}\:\tfrac{1}{2}\,\lambda^2\:{+}\:\tfrac{1}{6}\,\lambda^3\:{+}\:\dots\right)v\;{=}\;\exp^\lambda\,v \nonumber
		 %\end{align}
		 %\par It follows that $\lambda^\star\;{=}\;\exp^\lambda$. Combining with the required $\max\left\vert\lambda^\star\right\vert\;{<}\;1$ proves that the series $\sum_{p\,{\in}\,\mathbb{N}}\exp^{Ap}$ converges if and only if ${\nexists}\;\lambda(A)\:{\geq}\:0$. 
	\end{proof}
\end{lemma}

\begin{lemma}[Closed-loop reset instant]
	\par A RCS reset instant at a descending zero crossing, $t_\rho^\downarrow$, can, for any RC satisfying \thmref{thm:hosidf}, be computed using Assumptions \shortref{ass:xtr} to \shortref{ass:3}:
	\begin{align}
	\omega\,t_\rho^\downarrow\:{+}\:\angle\left(K(j\omega)\,S_L(j\omega)\,R_I(j\omega)\right)\:{+}\:\Phi(\omega,\vec{x}(t_{r}^\downarrow))\;{=}\;\pi \label{eq:restimes}
	\end{align}
	\par This $t_\rho^\downarrow$ corresponds to the zero crossings of $Q^\star$ with:
	\begin{align}
	{\angle}\:Q^\star(\omega)\;{=}\;\angle\left(K(j\omega)\,S_L(j\omega)\,R_I(j\omega)\right)\:{+}\:\Phi(\omega,\vec{x}(t_{r}^\downarrow)) \label{eq:phaseQ}
	\end{align}
	\par If $\Phi(\omega,\vec{x}(t_{r}^\downarrow))\;{=}\;0$, \eqref{eq:restimes} computes the descending zero crossings of the BLS. Phase shift $\Phi(\omega,\vec{x}(t_{r}^\downarrow))$ is defined as:
	\begin{multline} \Phi(\omega,\vec{x}(t_{r}^\downarrow))\;{\triangleq}\;\sin^{-1}\left(C_Q\,\sum\nolimits_{p\:{\in}\:2\mathbb{N}}\,\left(\exp^{A_Q\frac{p\pi}{\omega}}\:{-}\:\exp^{A_Q\frac{(p-1)\pi}{\omega}}\right)\right.\\ \left.{\times}\:B_Q\,\vec{x}(t_{\rho})\,\left(\left\vert K(j\omega)\,S_L(j\omega)\,R_I(j\omega)\right\vert\right)^{-1} \vphantom{\sum\nolimits_{p\:{\in}\:2\mathbb{N}}}\right) \label{eq:phi}
	\end{multline}
	where $A_Q,\,B_Q,\,C_Q$ and $D_Q$ denoting state-space matrices of $Q_\delta(s)\;{=}\;K(s)\,S_L(s)\,G(s)\,R_\delta(s)$. \lemref{lem:convergence} states series convergence, which requires asymptotic stability of $Q_\delta$. Set $\mathbb{N}$ is taken to exclude zero throughout this work.
	\begin{proof}
		\par Take \eqref{eq:E} for a SISO RC and pre-multiply by $K(s)$ to acquire the closed-loop description of $Q(s)$:
		\begin{multline}
		Q(s)\;{=}\;K(s)\,S_L(s)\,R_I(s)\\ \:{-}\:K(s)S_L(s)\,G(s)\, R_{\delta}(s)\sum_{t_r\in t_{R}}\vec{x}(t_{r})\,\exp^{{-}\:t_{r}\, s} \nonumber
		\end{multline}
		\par Combining \assref{ass:xtr} with \thmref{thm:per} shows that $t_R\;{=}\;t_\rho\:{+}\:p\,\pi\,{/}\,\omega,\ p\:{\in}\:\mathbb{Z}$, where $t_\rho$ has one entry. Inserting this and substituting $Q_\delta(s)$ as defined above gives:
		\begin{multline}
		Q(s)\;{=}\;K(s)\,S_L(s)\,R_I(s)\\ \:{-}\:Q_\delta(s)\sum_{p\:{\in}\:\mathbb{Z}}\vec{x}(t_{\rho}\,{+}\,\frac{p\,\pi}{\omega})\,\exp^{{-}\:(t_{\rho}\,{+}\,\frac{p\,\pi}{\omega})\, s} \nonumber
		\end{multline}
		\par Reset instant periodicity causes all resets instants prior to $t_r$ to be at times $t_r\:{-}\:\frac{p\,\pi}{\omega},\ p\:{\in}\:\mathbb{N}$. The time-domain solution for $\vec{q}(t)$ is obtained for a sinusoidal $r_I$ with frequency $\omega$:
		\begin{multline}
		\vec{q}(t_r)\;{=}\;\left\vert K(j\omega)\,S_L(j\omega)\,R_I(j\omega)\right\vert\,\\ {\times}\:\sin\left(\omega t\:{+}\:\angle\left(K(j\omega)\,S_L(j\omega)\,R_I(j\omega)\right)\right)\\
		\:{-}\:\sum_{p\:{\in}\:\mathbb{N}}\left(C_Q\,\exp^{A_Q\frac{p\pi}{\omega}}\,B_Q\:{+}\:D_Q\right)\,\vec{x}(t_{\rho}\,{+}\,\frac{p\,\pi}{\omega})\nonumber
		\end{multline}
		where the impulse response is expressed in state-space terms. \corref{cor:laplace} shows that $R_\delta(s)$ has no direct feed-through, which by definition of $Q_\delta(s)$ implies that $D_Q\;{=}\;0$. \\
		The following expression is obtained by utilizing the periodicity of $\vec{x}(t)$ proven by \thmref{thm:per}:
		\begin{multline}
		\vec{q}(t_r)\;{=}\;\left\vert K(j\omega)\,S_L(j\omega)\,R_I(j\omega)\right\vert\,\\{\times}\:\sin\left(\omega t_r\:{+}\:\angle\left(K(j\omega)\,S_L(j\omega)\,R_I(j\omega)\right)\right)\\
		\:{+}\:C_Q\left(\sum\nolimits_{p\:{\in}\:2\mathbb{N}}\,\exp^{A_Q\frac{p\pi}{\omega}}\:{-}\:\exp^{A_Q\frac{(p-1)\pi}{\omega}}\right)\,B_Q\,\vec{x}(t_{r})\nonumber
		\end{multline}
		\par Per definition of a zero-crossing reset law $\vec{q}(t_r)\;{=}\;0$. Inserting this and taking the inverse sine gives, for $m\:{\in}\:\mathbb{Z}$:
		\begin{align}	
		m\,\pi\;&{=}\;\begin{multlined}[t]\omega\, t_r\:{+}\:\angle\left(K(j\omega)\,S_L(j\omega)\,R_I(j\omega)\right)\\
		{+}\:\sin^{-1}\left[C_Q\,\sum\nolimits_{p\:{\in}\:2\mathbb{N}}\,\left(\exp^{A_Q\frac{p\pi}{\omega}}\:{-}\:\exp^{A_Q\frac{(p-1)\pi}{\omega}}\right)\right.\,\\
		\left.{\times}\:B_Q\,\vec{x}(t_{r})\,\left(\left\vert K(j\omega)\,S_L(j\omega)\,R_I(j\omega)\right\vert\right)^{-1}\vphantom{\exp^{A_Q\frac{p\pi}{\omega}}}\right] \end{multlined}\nonumber
		\end{align}
		\par The inverse sine exists if \assref{ass:3} holds. A descending zero-crossing occurs if the sinusoid argument is $m\,\pi$, with odd $m$. These cases correspond to $t\;{=}\;t_r^\downarrow$. The zero crossing direction is assumed to be unaffected by prior resets by \assref{ass:rup}. The solution $m\:{=}\:1$ is chosen. Inserting that while substituting $\Phi(\omega,\vec{x}(t_{r}^\downarrow))$ yields \eqref{eq:restimes}. Zero-crossings of a sinusoidal $Q^\star$ with phase \eqref{eq:phaseQ} gives resets $t_\rho^\downarrow\:{+}\:p\,\pi\,{/}\,\omega$.
	\end{proof}
\end{lemma}

\begin{assumption}[Small effect of resets on reset times] \label{ass:2}
	\par Assume that $\Phi(\omega,\vec{x}(t_{\rho}))$ \eqref{eq:phi} satisfies $\Phi(\omega,\vec{x}(t_{\rho}))\:{\ll}\:\pi,\ {\forall}\:\omega$. This holds if reset times are close to the BLS zero crossings of $\vec{q}(t)$.
\end{assumption}

\begin{lemma}[States $\vec{x}(t_\rho^\downarrow)$] \label{lem:x}
	\par States $\vec{x}(t_\rho^\downarrow)$ are, given Assumptions \shortref{ass:xtr} to \shortref{ass:2}, for a system satisfying \thmref{thm:per}, computed by:
	\begin{multline}
		\vec{x}(t_\rho^\downarrow)\:{=}\:\left[\vphantom{\exp^{A_H\,\frac{(k-1)\pi}{\omega}}}I\,{+}\,\left(\omega^2\,I^2\,{+}\,A_R^2\right)^{-1}A_R\,B_R\,C_Q\,\sum\nolimits_{k\:{\in}\:2\mathbb{N}}\,\right.\\
		\left(\exp^{A_Q\frac{k\pi}{\omega}}\:{-}\:\exp^{A_Q\frac{(k-1)\pi}{\omega}}\right)\,B_Q\:{-}\:0.5\,C_H\sum\nolimits_{k\:{\in}\:2\mathbb{N}}\\
		\left.\left(\exp^{A_H\,\frac{k\pi}{\omega}}\:{-}\:\exp^{A_H\,\frac{(k-1)\pi}{\omega}}\right)\,B_H\right]^{-1}\left(\omega^2\,I\:{+}\:A_R^2\right)^{-1}\omega\,B_R\\ 
		{\times}\:\left\vert K(j\omega)\,S_L(j\omega)\,R_I(j\omega)\right\vert \label{eq:result}
	\end{multline}
	\par Which requires the inverted terms to be invertible. State-space matrices $A_H,\,B_H,\,C_H$ and $D_H$ correspond to $H(s)$:
	\begin{align}
	H(s)\;&{=}\;R^X_{\delta}(s)\:{-}\:R^X_{L}(s)\,K(s)\,S_L(s)\,G(s)\,R_{\delta}(s) \nonumber
	\end{align}
	\par The two series converge if $A_Q$ and $A_H$ satisfy \lemref{lem:convergence}, thus if $Q$ and $H$ are asymptotically stable.
	\begin{proof}
		\par Consider \eqref{eq:X} for a SISO RC and separate it into $X_L(s)$ and $X_\delta(s)$ such that $X(s)\;{=}\;X_L(s)\:{+}\:X_\delta(s)$:
		\begin{align}
		X_L(s)\;&{=}\;R^X_{L}(s)\,K(s)\,S_L(s)\,R_I(s) \nonumber\\
		X_\delta(s)\;&{=}\;\begin{multlined}[t]\left(R^X_{\delta}(s)\:{-}\:R^X_{L}(s)\,K(s)\,S_L(s)\,G(s)\,R_{\delta}(s)\right)\\
		{\times}\,\sum_{t_r\in t_{R}}\vec{x}(t_r\in t_{r})\,\exp^{{-}\:t_{r,}\,s}\end{multlined}\nonumber
		\end{align}
		\par Writing the linear term $\vec{x}_L(t)$ in time domain for a sinusoidal input, while using that $\vec{x}_L(t)$ is real, gives the following:
		\begin{multline}
		\vec{x}_L(t)\;{=}\;\Re\left\{\vphantom{\exp^{j\omega (t-\frac{\pi}{2})}}R^X_L(j\omega)\,\left\vert K(j\omega)\,S_L(j\omega)\,R_I(j\omega)\right\vert\right.\\ \left.{\times}\:\exp^{j\omega (t-\frac{\pi}{2})\:{+}\:j\angle\left(K(j\omega)\,S_L(j\omega)\,R_I(j\omega)\right)}\right\} \nonumber
		\end{multline}
		\par Evaluating at $t_\rho^\downarrow$ by rewriting and inserting \eqref{eq:restimes} shows:
		\begin{multline}
		\vec{x}_L(t_\rho^\downarrow)\;{=}\;\Re\left\{\vphantom{\exp^{j\omega (t-\frac{\pi}{2})}}R^X_L(j\omega)\,\left\vert K(j\omega)\,S_L(j\omega)\,R_I(j\omega)\right\vert\right.\\ \left.{\times}\:\exp^{j\left(\frac{\pi}{2}\:{-}\:\Phi(\omega,\vec{x}(t_{\rho^\downarrow}))\right)}\right\} \nonumber
		\end{multline}
		\par Take $\Phi(\omega,\vec{x}(t_{\rho}^\downarrow))$ from \eqref{eq:phi} and apply \assref{ass:2}, such that $\Phi(\omega,\vec{x}(t_{\rho}^\downarrow))\:{=}\:\sin^{-1}\left(\bullet\right)\;{\approx}\;\left(\bullet\right)$, where $\bullet$ denotes the terms within $\Phi$. Then, the last term of $\vec{x}_L(t_\rho^\downarrow)$ becomes $\exp^{j\left(\frac{\pi}{2}-\bullet\right)}$. Given \assref{ass:2}, the first Taylor expansion can be used, giving the simplification $\exp^{j\left(\frac{\pi}{2}-\bullet\right)}\;{\approx}\;j\:{+}\:\left(\bullet\right)$. Inserting this and expanding $\Phi(\omega,\vec{x}(t_{\rho}^\downarrow))$ gives:
		\begin{align}
		\vec{x}_L(t_\rho^\downarrow)\;{=}\;\begin{multlined}[t]\Re\left\{\vphantom{\exp^{A_Q\frac{(k-1)\pi}{\omega}}}R^X_L(j\omega)\,\left\vert K(j\omega)\,S_L(j\omega)\,R_I(j\omega)\right\vert\right.\\ \left.{\times}\:\left[j\:{+}\:	C_Q\,\sum\nolimits_{p\:{\in}\:2\mathbb{N}}\,\left(\exp^{A_Q\frac{p\pi}{\omega}}\:{-}\:\exp^{A_Q\frac{(p-1)\pi}{\omega}}\right)\right.\right.\\ \left.\left.{\times}\:B_Q\,\vec{x}(t_{\rho}^\downarrow)\,\left(\left\vert K(j\omega)\,S_L(j\omega)\,R_I(j\omega)\right\vert\right)^{-1}\vphantom{\exp^{A_Q\frac{(k-1)\pi}{\omega}}}\right]\right\} \end{multlined} \nonumber
		\end{align}
		\par For a SISO RC this can be simplified to:
		\begin{align}
		\vec{x}_L(t_\rho^\downarrow)\;{=}\;\begin{multlined}[t]\Re\left\{\vphantom{\exp^{A_Q\frac{(p-1)\pi}{\omega}}}R^X_L(j\omega)\,\left\vert K(j\omega)\,S_L(j\omega)\,R_I(j\omega)\right\vert\,j\right.\\ {+}\:R_L^X(j\omega)\,C_Q\,\sum\nolimits_{p\:{\in}\:2\mathbb{N}}\,\left(\exp^{A_Q\frac{p\pi}{\omega}}\right.\\
		\left.\left.{-}\:\exp^{A_Q\frac{(p-1)\pi}{\omega}}\right)\,B_Q\,\vec{x}(t_{\rho}^\downarrow)\right\} \end{multlined} \nonumber
		\end{align}
		\par Consider $X_\delta(s)$. Insert $H(s)$ and write in time domain. As $R_\delta(s)$ has no direct feed-through $D_H$ must equal zero.
		\begin{align}
		\vec{x}_\delta(t)\;&{=}\;\sum_{t_r\in t_{{R}_{\leq t}}}\left(C_H\,\exp^{A_H(t-t_r)}\,B_H\right)\,\vec{x}(t_{r})\nonumber
		\end{align}
		\par Considering \assref{ass:xtr} with \thmref{thm:per}, such that all resets are spaced $\pi\,{/}\,\omega$ apart, and evaluating at $t_\rho^\downarrow$ gives:
		\begin{align}
		\vec{x}_\delta(t_\rho^\downarrow)\;&{=}\;C_H\sum_{p\:{\in}\:2\mathbb{N}}\left(\exp^{A_H(\frac{p\pi}{\omega})}-\exp^{A_H(\frac{(p-1)\pi}{\omega})}\right)\,B_H\,\vec{x}(t_{\rho}^\downarrow)\nonumber
		\end{align}
		\par Inserting these results into  $\vec{x}(t_\rho^\downarrow)\;{=}\;\vec{x}_L(t_\rho^\downarrow)\:{+}\:\vec{x}_\delta(t_\rho^\downarrow)$, solving for $\vec{x}(t_\rho^\downarrow)$ and inserting \eqref{eq:RLXj}, \eqref{eq:RLX} gives the stated result.
	\end{proof}
\end{lemma}

\begin{theorem}[Analytical solution for $E(\omega)$ ($\delta$-CL)]\label{thm:E}
	The error response of a RCS \eqref{eq:CL} satisfying \thmref{thm:per}, given Assumptions \shortref{ass:xtr} to \shortref{ass:2}, is stated below. Arguments of $B^\star(\omega,\vec{x}(t^\downarrow_{r}))$ are dropped.
	\begin{align}
	&E_{\delta{-}CL,n}(\omega)\;{=}\;S_L(\omega n)\nonumber\\&{\times}\begin{cases}
	R_I(\omega)\:{-}\:\,G(\omega)\,R_{DF,1}^\star(\omega,B^\star)\,\Psi(\omega,1),&n\;{=}\;1\\
	{-}\:G(\omega n)R_{DF,n}^\star(\omega,B^\star)\,\Psi(\omega,n),\, &n\;{>}\;1\end{cases}\label{eq:Eresult}  
	\\
	&\text{where:}\\&\quad \Psi(\omega,n)\;{=}\;\begin{multlined}[t]\left(\vphantom{\exp^{nj\angle K(\omega)}}\left\vert K(\omega)\,S_L(\omega)\,R_I(\omega)\right\vert\right.\\ \\\left.{\times}\:\exp^{nj\angle K(\omega)\,S_L(\omega)\,R_I(\omega)\:{+}\:nj\Phi(\omega,\vec{x}(t_r^\downarrow))}\right) \end{multlined}\nonumber
	\end{align}	
	\par \defref{def:hosidf2} defines $R^\star_{DF,n}$. Parameters $\vec{x}(t^\downarrow_{r})$, $\Phi(\omega,\vec{x}(t_r^\downarrow))$ and $B^\star(\omega,\vec{x}(t^\downarrow_{r}))$ are given by \eqref{eq:result}, \eqref{eq:phi} and \eqref{eq:qdf}, respectively.
	\begin{proof}
		\par Insert \eqref{eq:result} into \eqref{eq:qdf} to solve \eqref{eq:Ehosidf}. Virtual input $Q^\star$ has magnitude $1$, see \corref{lem:bstar}, and phase \eqref{eq:phaseQ}. From \assref{ass:xtr} it follows that a summation over $t_{\rho}$ is obsolete, as this set has one entry only. Rewriting gives the result.
		\par These equations combine harmonic and reference frequencies. Multiplying the phase by $n$ accounts for this.
	\end{proof}
\end{theorem}

\begin{corollary}[Analytical solution for $S(\omega)$]\label{cor:S}
	\par Sensitivity is defined as $S(\omega)\;{=}\;E(\omega)\,R_I(\omega)^{-1}$. Applying this to \eqref{eq:Eresult} while correcting the phase for harmonics gives:
	\begin{align}
	S_{\delta{-}CL,n}(\omega)\;{=}\;E_{\delta{-}CL,n}(\omega)\,\left(\vert R_I(\omega)\vert\,\exp^{nj\angle R_I(\omega)}\right)^{-1} \label{eq:Sresult}
	\end{align}
\end{corollary}

\begin{corollary}[Complementary Sensitivity]
	\par The complementary sensitivity $T(\omega)$ is defined as $I\:{-}\:S(\omega)$:
	\begin{align}
	T_{\delta{-}CL,n}(\omega)\;{=}\;I\:{-}\:E_{\delta{-}CL,n}(\omega)\,\left(\vert R_I(\omega)\vert\,\exp^{nj\angle R_I(\omega)}\right) \label{eq:T}
	\end{align}
\end{corollary}

\begin{corollary}[Control Sensitivity]
	\par Split the linear system $G(s)$ in plant $P(s)$ and controller $C(s)$, $G(s)\;{=}\;P(S)\,C(S)$. Control input $U(s)$ enters $P(s)$: $Y(s)\:{=}\:P(s)\,U(s)$. The control input\\ $CS(\omega)\:{:}\:\vec{r}_I(t)\:{\mapsto}\:\vec{u}(t)\;{\triangleq}\;CS(\omega)\;{=}\;P^{-1}(\omega)\,T(\omega)$:
	\begin{align}
	CS_{\delta{-}CL,n}(\omega)\;{=}\;P^{-1}(jn\omega)\,T_{\delta{-}CL,n}(\omega) \label{eq:CS}
	\end{align}
\end{corollary}

\par These results can be transformed into time-domain signals:
\begin{align}
\vec{y}(t)\;&{\approx}\;\sum\nolimits_{n=1}^\infty \left\vert T_{DF,n}(\omega)\right\vert\sin(n\omega t\:{+}\:\angle T_{DF,n}(\omega))\\
\vec{e}(t)\;&{\approx}\;\sum\nolimits_{n=1}^\infty \left\vert E_{DF,n}(\omega)\right\vert\sin(n\omega t\:{+}\:\angle E_{DF,n}(\omega))\\
\vec{u}(t)\;&{\approx}\;\sum\nolimits_{n=1}^\infty \left\vert CS_{DF,n}(\omega)\right\vert\sin(n\omega t\:{+}\:\angle CS_{DF,n}(\omega))
\end{align}

\begin{assumption}[Superposition for multi-sine inputs] \label{ass:5}
	\par Consider a closed-loop RC \eqref{eq:CL} with multiple sinusoidal references superimposed, having magnitudes $R_{I_1}$ to $R_{I_{\bar{k}}}$, $\bar{k}\,{\in}\,\mathbb{N}$. Define the corresponding BLS magnitudes for $\vec{q}(t)$ as ${\vert} \vec{q}_{I_1}{\vert}$ to $\vert \vec{q}_{I_{\bar{k}}}\vert$, individually computed for each reference. If ${\exists}\:p\:{:}\:{\vert}\vec{q}_{I_k}{\vert}\:{\gg}\:{\vert}\vec{q}_{I_j}{\vert},\:{\forall}\:j\:{\in}\:\left\{1,\dots,\bar{k}\right\},\:j\:{\neq}\:k$, assume that solely reference $R_{I_k}$ determines reset times and weights. If so, all other references are handled by the BLS and can be merged with the nonlinear RC response for $R_{I_{k}}$ through superposition. This allows modelling of multi-sine references \cite{Saikumar2020}.
	\par This framework extends to permitting disturbances, as any sinusoidal disturbance $\vec{d}$ after the nonlinear reset element gives, for a linear plant, some sinusoidal signal $\vec{q}_d$. This can be handled analogous to $\vec{q}_{I_j}$ as described above.
\end{assumption}

\begin{figure}[b]
	\centering
	\includegraphics[width=\linewidth]{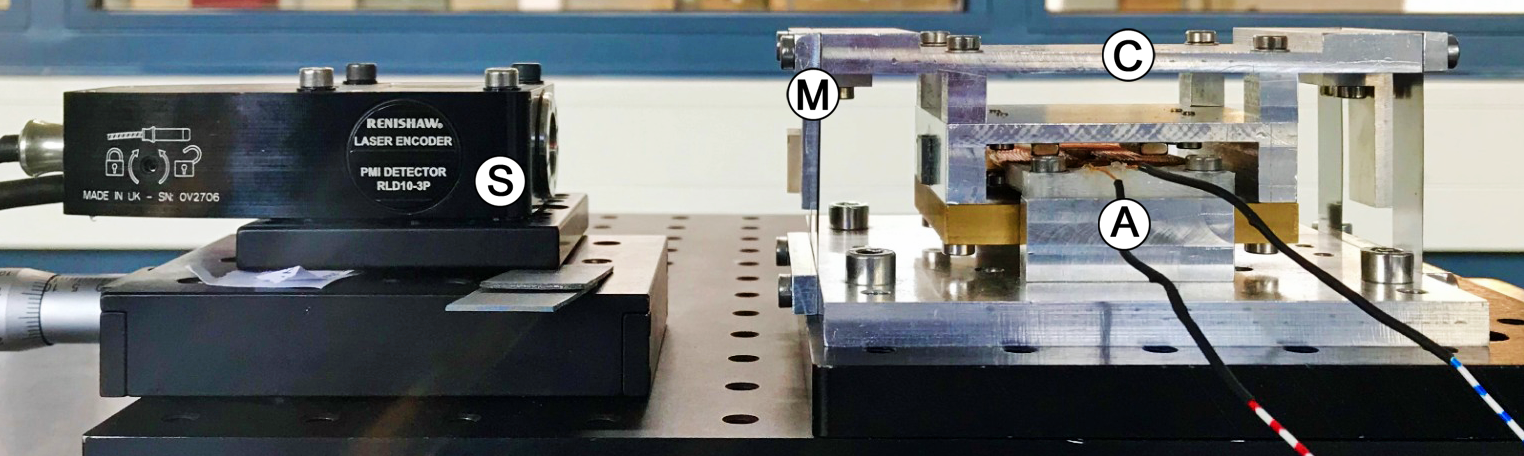}
	\caption{1 Degree-of-Freedom precision positioning stage that moves cart (C) using Lorentz actuator (A). This cart is attached to the frame by means of two leaf flexures, which constrain all movements but one translation. This translation is measured using laser encoder (S), which measures its distance relative to cart-fixed mirror (M).}
	\label{fig:setup} 
\end{figure}
\section{Setup}
\label{sec:setup}

\par The implications of the various assumptions are investigated in this section, evaluating where they cause $\delta$-CL to not predict $\vec{e}(t)$ correctly. First, a precision positioning system will be introduced. Afterwards, performance metrics are defined.
\par For analysis $n\;{=}\;1000$ harmonics are used. Series over impulse responses as in \eqref{eq:phi}, \eqref{eq:result} are evaluated with sufficient terms to ensure convergence.

\subsection{Precision Positioning Stage}
\par The 1 Degree-of-Freedom precision positioning stage depicted in \figref{fig:setup} is used to validate the derived method. This SISO stage is a classic mass-spring-damper system. The transfer function of this stage is identified to be:
\begin{align}
P(s)\;{=}\;\frac{3.038\:{\times}\:10^4}{s^2\:{+}\:0.7413\,s\:{+}\:243.3} \label{eq:setup}
\end{align}
\par
\begin{figure}[t]
	\centering
	\includegraphics[width=\linewidth]{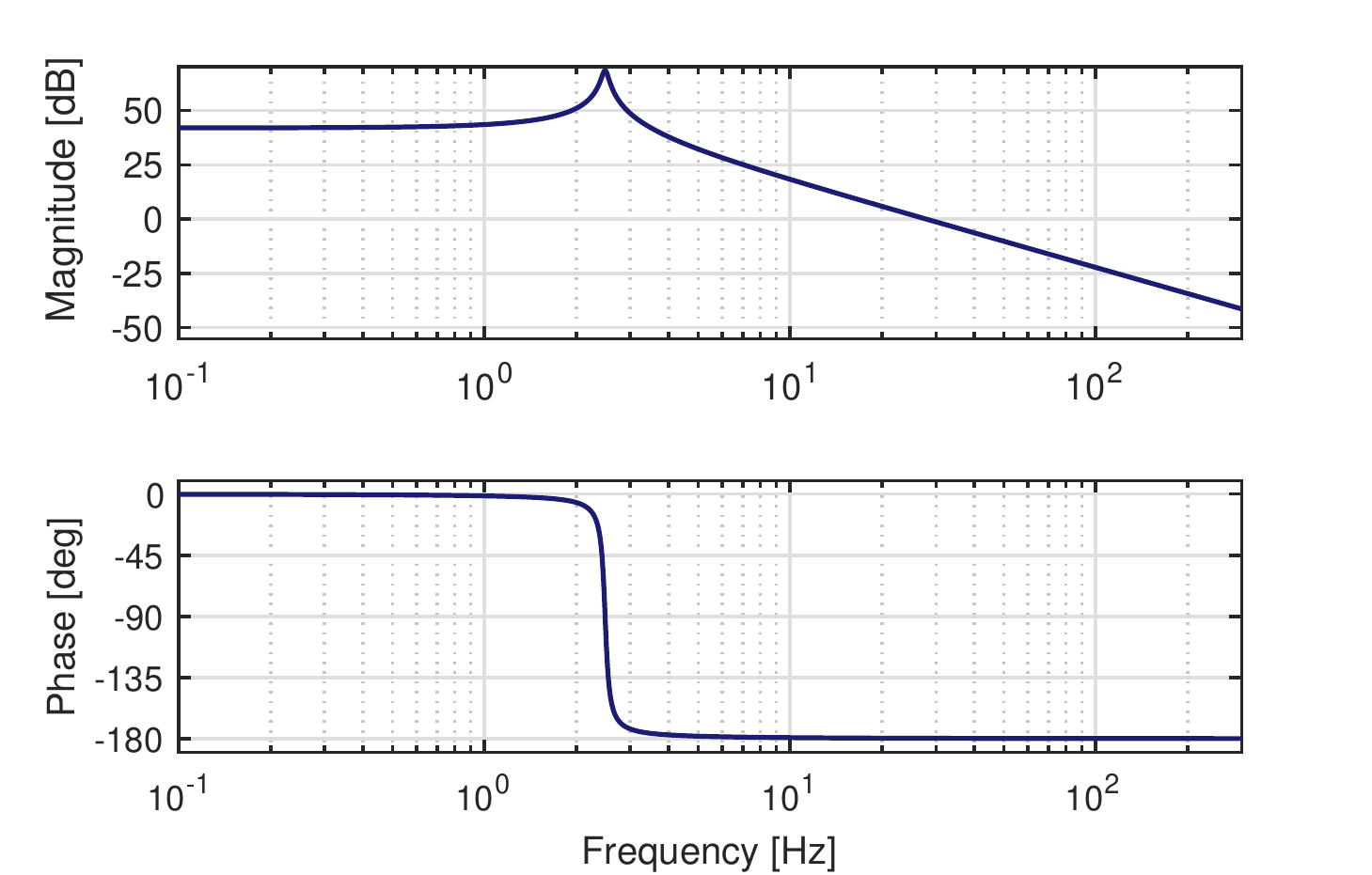}
	\caption{Bode Plot corresponding to the system \eqref{eq:setup} shown in \figref{fig:setup}.}
	\label{fig:bodesetup} 
\end{figure}

 The corresponding Bode Plot is given in \figref{fig:bodesetup}. For the sake of analysis consider the simple case where (i) there is no noise, (ii) there are no disturbances, (iii) no quantization effects are present and (iv) a continuous-time controller is used.
\subsection{Controllers}
\par A linear PID controller $C(s)$ is added between the reset element and the plant, such that $G(s)\;{=}\;P(s)\,C(s)$. Parameter $\beta$ is introduced, placing zero $\omega_d$ and pole $\omega_t$ symmetrically around bandwidth, defined as crossover frequency $\omega_c$.
\begin{align}
C(s)\;&{=}\;k_p\,{\left(\frac{s\:{+}\:\omega_i}{s}\right)}\,{\left(\frac{s\:{+}\:\omega_c\,{/}\,\beta}{s\:{+}\:\omega_c\,\beta}\right)},\quad \beta\;{=}\;\frac{\omega_c}{\omega_d}\;{=}\;\frac{\omega_t}{\omega_c}
\end{align}
\par Bandwidth $\omega_c$ defined as the gain cross-over frequency is set to $100$ Hz. Gain $k_p$ is adjusted to achieve this bandwidth, based on DF analysis. For all implementations, $\omega_i\;{=}\;10$ Hz is chosen.
\par Various CgLp-PID controller combinations tuned for different specifications are used for validation. \tabref{tab:testtuning} provides the corresponding tuning parameters. $PM_{BLS}$ denotes the BLS phase margin. Let $PM_{DF}$ be the phase margin as predicted by DF analysis. Then, the phase added through RC is given by $\phi_{RC}\;{=}\;PM_{DF}\:{-}\:PM_{BLS}$.
\begin{table}[t]
	\centering
	\caption{Parameters for various CgLp and PID controller designs. For all $\mathcal{R}^\star$ controllers $\omega_i\;{=}\;10$ Hz, $\omega_c\;{=}\;100$ Hz and $\omega_f\;{=}\;500$ Hz. Gain $k_p$ is adjusted to achieve bandwidth $\omega_c$.}
	\label{tab:testtuning}
	\begin{tabular}{lcccccc}
		\hline
		& & & & & & \\  [-0.8em]
		& $PM_{BLS}$ & $\phi_{RC}$ & $\gamma$ & $\omega_r$ [Hz] & $\alpha$ & $\beta$ \\ \hline
		& & & & & & \\  [-0.8em]
		${\mathcal{R}^\star_0}$ & $30^\circ$ & $20^\circ$ & $0$        	& $98.38$ & $1.07$ & $2.67$  \\
		${\mathcal{R}^\star_1}$ & $30^\circ$ & $20^\circ$ & $0.5$ 		& $23.08$ & $1.04$ & $2.57$  \\
		${\mathcal{R}^\star_2}$ & $20^\circ$ & $20^\circ$ & $0.5$ 		& $23.08$ & $1.04$ & $2.03$  \\
		%${\mathcal{CI}^\star_0}$ & $20^\circ$ & $20^\circ$ & $0$ 	 	&  		  & 	   & $3.73$  \\
		\hline	
	\end{tabular}
\end{table}
\subsection{Performance Metrics}
\par The signal $\vec{e}(t)$ as predicted by \eqref{eq:Eresult} is compared to the corresponding simulated signal. A metric often used in literature for capturing the time-domain prediction accuracy is Integral Square Error (ISE). A normalized version is given by:
\begin{align}
ISE(\omega)\;&{\triangleq}\;\frac{\int (\vec{e}_\omega(t)\:{-}\:\hat{\vec{e}}_\omega(t)^2\,\textrm{d}t}{\int \hat{\vec{e}}_\omega\!\!^2(t)\,\textrm{d}t}
\end{align}
where simulation data is denoted by $\vec{e}$ and prediction data by $\hat{\vec{e}}$. A time vector with parameter $\omega$, such as $\vec{e}_\omega(t)$, denotes the time response for a reference with frequency $\omega$.
\par In the high-tech industry peak error values indicate precision, described by the $\mathcal{L}_\infty$ norm, which is normalized:
\begin{align}
L_\infty(\omega)\;{\triangleq}\;\frac{\vert \max_t \vert \vec{e}_\omega(t)\vert\:{-}\:\max_t \vert \hat{\vec{e}}_\omega(t)\vert\vert}{\max_t \vert \hat{\vec{e}}_\omega(t)\vert}
\end{align}

%\par The last metric used is the peak sensitivity value, $S_\infty(\omega)$. It is defined for a sinusoidal reference $\vec{r}_I(t)$ with amplitude $r_0$. Let $\vec{e}(t)$ be the output to such a reference. Then:
%\begin{align}
%S_\infty(\omega)\;{=}\;\frac{\max_t\vert\vec{e}(t)\vert}{r_0}
%\end{align}
\vspace{-1mm}
\section{Validation}
\label{sec:val}
\par \tabref{tab:assumptions} provides an overview of the various assumptions used by the three analytical RCS describing methods. CL-DF and $\delta$-CL use similar assumptions, except for using different reset positions and CL-DF assuming a sinusoidal $\vec{q}(t)$. 
\par Assumptions \shortref{ass:rup} and \shortref{ass:3} of $\delta$-CL are not mentioned, because no results in this paper found \assref{ass:rup} to not hold, whilst \assref{ass:3} is violated only for a few frequencies in one result, \figref{fig:pm2}. \assref{ass:5} is used to alleviate the constraint on having a sinusoidal $\vec{r}_I(t)$. This case will be demonstrated, after the effects of other assumptions on sensitivity prediction errors are verified using a sinusoidal input.
\begin{table}[t]
	\centering
	\caption{Overview of assumptions analytical methods for computing frequency-domain closed-loop RC behaviour use. Empty fields indicate that there are no assumptions. Note that assumptions on $\vec{r}_I$ do not have to cause errors, as $\vec{r}_I$ is designable.}
	\label{tab:assumptions}
	\begin{tabular}{lccc}
		\hline 
		& & & \\[-0.8em]
		& DF   & CL-DF      & $\delta$-CL        \\ \hline
		& & & \\[-0.8em]
		Modelled resets per period:     & $2$           & $2$         &   $2$       \\
		Signals assumed sinusoidal:           & $\vec{q}(t)$  & $\vec{r}_I(t)$& $\vec{r}_I(t)$\\
		Resets assumed at:         &             & $\vec{q}_{DF,1}\;{=}\;0$  & $\vec{q}_{BLS}\;{\approx}\;0$ \\
		Neglects harmonics:         & Yes            &            &           \\ \hline
	\end{tabular}
\end{table}
\vspace{-1mm}
\subsection{Effects of \assref{ass:xtr}}
\par \assref{ass:xtr} simplifies analysis by modelling two resets per period only, an assumption used by all analytical methods under consideration. However, this is known to invoke errors. A time domain example is provided to illustrate the types of errors inflicted. Then, it is shown how time regularization can decouple this error source from other sources.
\vspace{0.5mm}
\par
\begin{figure}[b]
	\centering
	\includegraphics[width=\linewidth]{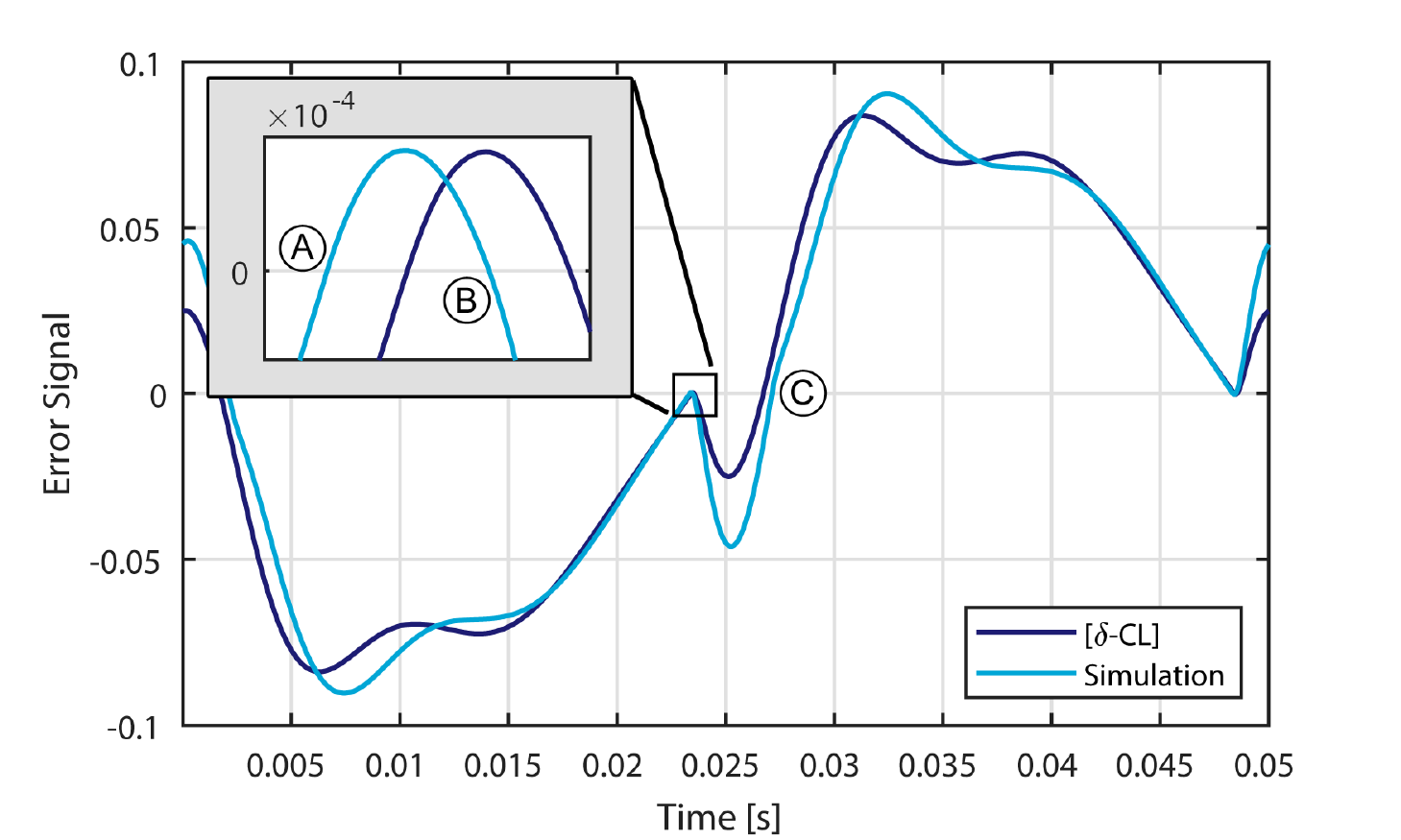}
	\caption{Simulated and $\delta$-CL predicted error response for a $20$ Hz reference signal on controller $\mathcal{R}_1^\star$ and plant \eqref{eq:setup}, thus using $\gamma\;{=}\;0.5$. Points (A), (B) and (C) indicate reset instants. (B) is an undesired consecutive reset to (A), (C) is an additional reset.}
	\label{fig:gammaTime} 
\end{figure}

 \figref{fig:gammaTime} shows the simulated and $\delta$-CL modelled error signals for a RCS generating three resets per half period, marked as (A), (B) and (C). Reset (A) is the one modelled by $\delta$-CL, whereas (B) and (C) are not modelled, causing errors.
\vspace{0.5mm}
\subsubsection{Consecutive resets}
\par A pair of resets is said to be consecutive if they occur close together temporally, relative to period $\pi\,{/}\,\omega$. Reset (B) in \figref{fig:gammaTime} is therefore consecutive to (A). Let the corresponding reset times be $t_{A}$ and $t_{B}$, $t_{B}\;{>}\;t_{A}$. An ODE solution is used to express $\vec{x}(t_{B})$ in terms of $\vec{x}(t^+_{A})$:
\begin{align}
\vec{x}(t_{B})\;{=}\;e^{A_R(t_{B}-t_{A})}x^+(t_{A})\:{+}\:\int_{t_{A}}^{t_{B}}e^{A_R(t-\tau)}B_R\,u(\tau)\,\text{d}\tau \nonumber
\end{align}
\par Consider the limit case for consecutive resets, $t_{B}\:{\to}\:t_{A}$. Insert \eqref{eq:ODE} to obtain $\vec{x}(t^+_{B})$ as a function of $\vec{x}(t_{A})$:
\begin{align}
\lim_{t_{B}\to t_{A}}\,\vec{x}(t_{B})\;&{=}\;\vec{x}^+(t_{A})\;{=}\;A_\rho\,\vec{x}(t_{A}) \nonumber \\
\lim_{t_{B}\to t_{A}}\,\vec{x}^+(t_{B})\;&{=}\;A_\rho\,\vec{x}^+(t_{A})\;=\;A_\rho^2\,\vec{x}(t_{A}) \nonumber
\end{align}
\par When comparing to \eqref{eq:ODE} it follows that, for $t_{r,B}\:{\to}\:t_{r,A}$, the response becomes equivalent to that obtained by having one reset with reset matrix $A_\rho^2$. As all analytical methods model only one reset with reset matrix $A_\rho$, errors occur if $A_\rho\;{\neq}\;A_\rho^2$. For diagonal $A_\rho$ parametrized by $\gamma$, such as in \eqref{eq:CgLp}, modelling errors thus occur if $\gamma\;{\neq}\;\gamma^2\;{\Leftrightarrow}\;\gamma\;{\notin}\;\{0,1\}$. Full reset therefore does not invoke errors here. \figref{fig:gammaTime} uses $\gamma\;{=}\;0.5$, meaning that the actual response to resets (A), (B) is almost equivalent to having one reset with $\gamma\;{=}\;0.5^2\;{=}\;0.25$. This increases the reset-induced impulse weight \eqref{eq:X}, explaining the higher than modelled reset spike after (B).
\begin{figure}[b]
	\vspace{-1mm}
	\centering
	\includegraphics[width=0.98\linewidth]{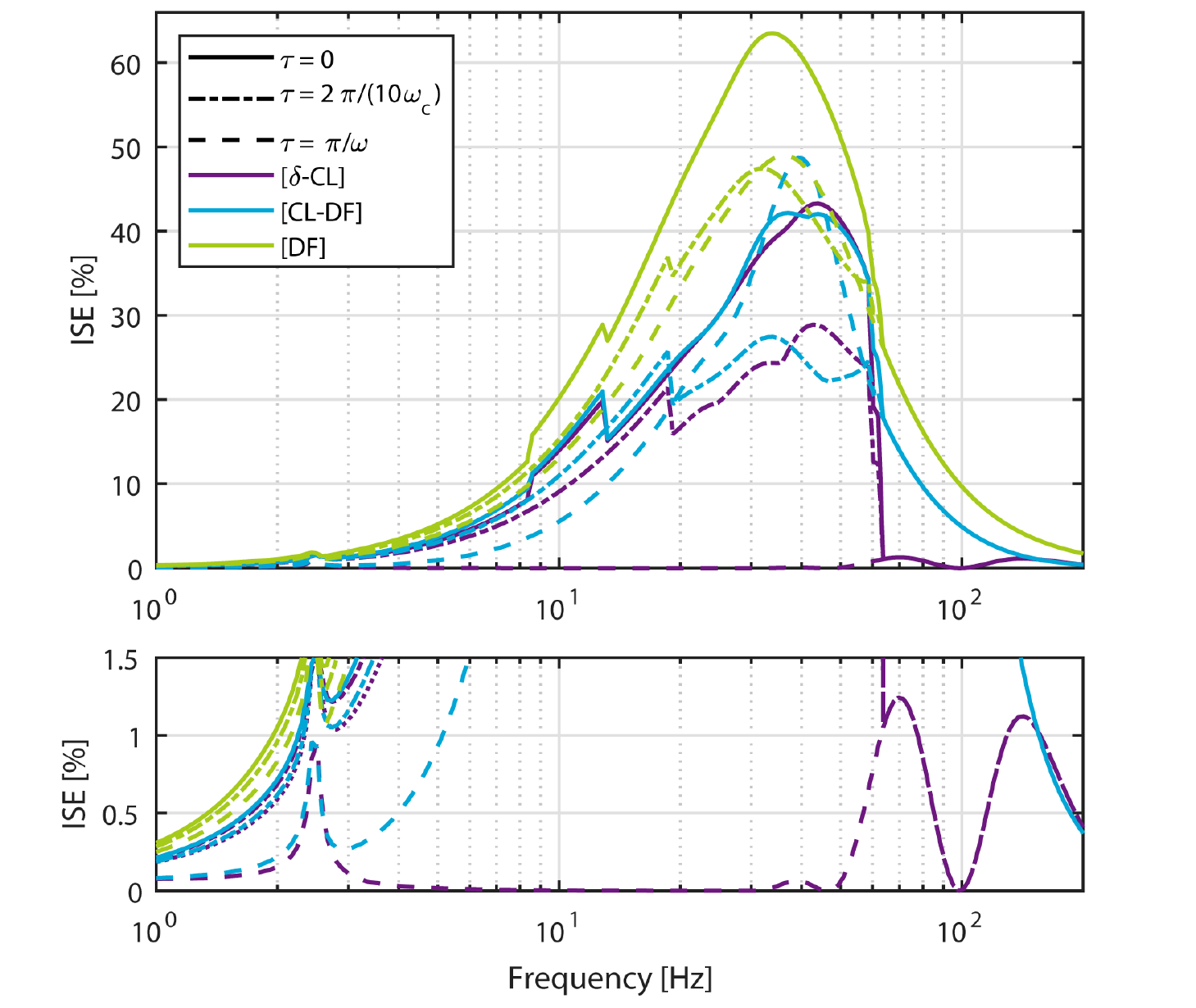}
	\caption{Normalized ISE values for the three prediction methods, using controller $\mathcal{R}^\star_1$ on \eqref{eq:setup}. Three different time regularization settings $\tau$ are used. The bottom figure provides a detail view on lower ISE values.}
	\label{fig:timereg} 
\end{figure}
\vspace{0.5mm}
\subsubsection{Additional resets}
\par Reset (C) in \figref{fig:gammaTime} is relatively far away from (A) in temporal sense. As such, $t_{C}\:{\to}\:t_{A}$ cannot be used here, implying that full reset is not exempt from errors caused by not modelling (C). \figref{fig:gammaTime} shows how (C) causes the error peak prediction to be wrong, inflicting $L_\infty$ errors.
\par
\begin{figure}[t]
	\centering
	\includegraphics[width=\linewidth]{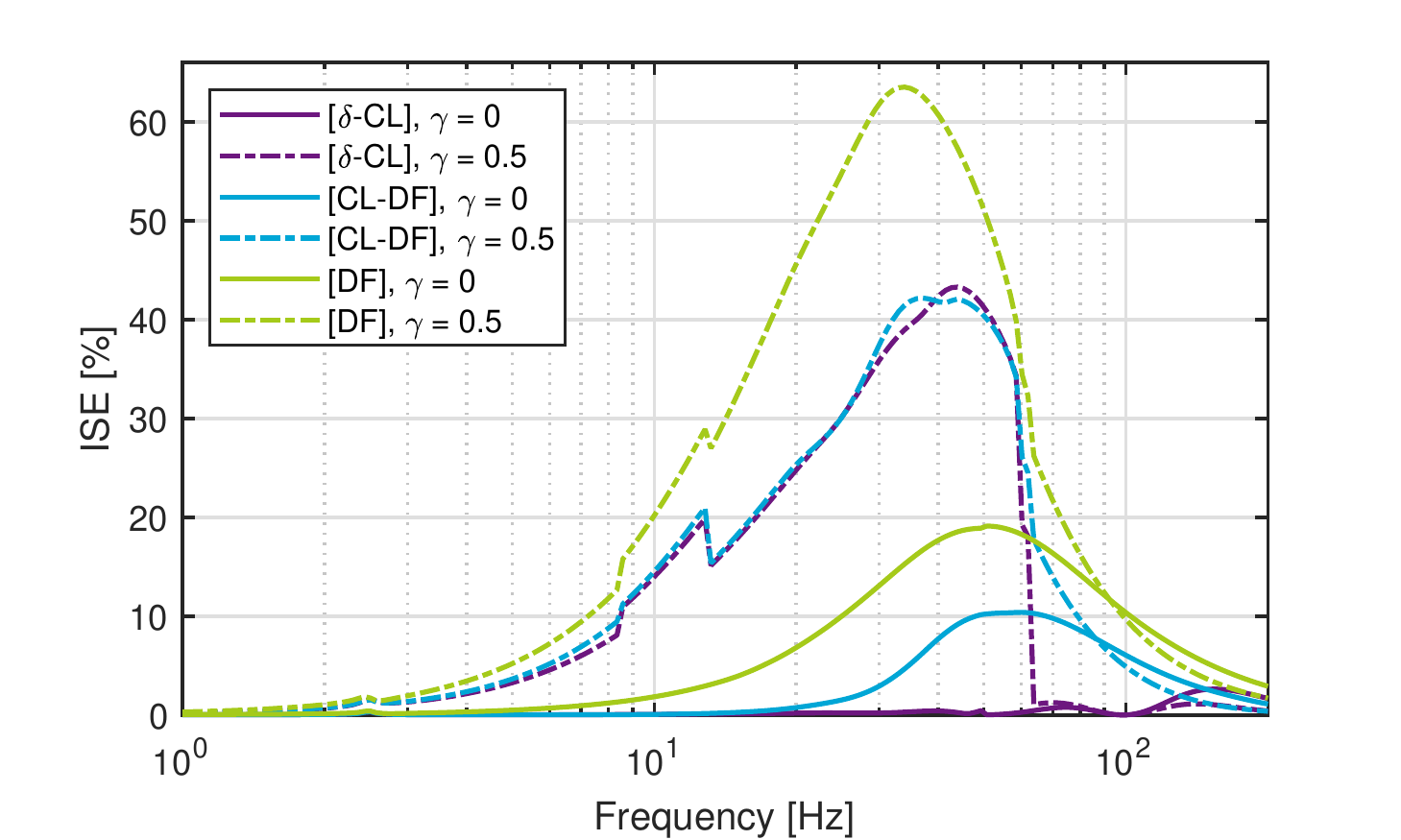}
	\caption{Normalized ISE values for the three analytical modelling methods. No time regularization is used. Results are provided for controllers $\mathcal{R}^\star_0$ ($\gamma\;{=}\;0$) and $\mathcal{R}^\star_1$ ($\gamma\;{=}\;0.5$), both controlling plant \eqref{eq:setup}.}
	\label{fig:gammaVals} 
\end{figure}

 \figref{fig:gammaVals} gives ISE results for all three methods, plotted for a range of reference frequencies. Controllers $\mathcal{R}^\star_0$ ($\gamma\;{=}\;0$) and $\mathcal{R}^\star_1$ ($\gamma\;{=}\;0.5$) are used to control \eqref{eq:setup}. Controller $\mathcal{R}^\star_0$ has, unlike $\mathcal{R}^\star_1$, negligible effects caused by consecutive resets because it is fully resetting. Both are affected by additional resets. Modelling errors of $\mathcal{R}^\star_1$ are, for all methods, at most frequencies, several factors above those of $\mathcal{R}^\star_0$, illustrating the considerable effects of consecutive resets on modelling accuracy. $\delta$-CL is seen to invoke comparatively small prediction errors when consecutive resets do not affect the response ($\gamma\;{=}\;0$).
\vspace{0.5mm}
\subsubsection{Time regularization}
\par Time regularization allows to remove consecutive or even additional resets, eliminating errors caused by \assref{ass:xtr}. The following options are used:
\begin{itemize}
	\item No time regularization: $\tau\;{=}\;0$, such that all errors caused by consecutive and additional resets remain visible.
	\item Optimal time regularization: $\tau\;{=}\;2\,\pi\,{/}\,(10\,\omega_c)$, $\omega_c$ in rad/s, suggested to be optimal when handling quantization \cite{Kieft}. This generally removes consecutive reset effects.
	\item Full time regularization: $\tau\;{=}\;\pi\,{/}\,\omega$, enforcing two resets per period, removing all errors caused by \assref{ass:xtr}.
\end{itemize}
\par Full time regularization does not necessarily improve RC system performance. However, it decouples modelling error sources for $\delta$-CL, as all remaining errors are caused by \assref{ass:2}. This is employed to simplify analysis.
\vspace{0.5mm}
\par All three time regularization methods are applied to partially resetting controller $\mathcal{R}^\star_1$. Corresponding ISE results are given in \figref{fig:timereg}. The DF and CL-DF descriptions show some improvement for more aggressive time regularization, but results are not consistent over all frequencies. $\delta$-CL shows a significant performance improvement for more aggressive time regularization. In case of full time regularization, ISE values for $\delta$-CL drop below $1.5\,\%$, errors that are thus solely caused by \assref{ass:2}. ISE values for CL-DF with full time regularization are caused by the assumed reset positions and sinusoidal $\vec{q}(t)$, see \tabref{tab:assumptions}. These two assumptions thus inflict considerably larger errors than \assref{ass:2} of $\delta$-CL.
\begin{figure}[t]
	\centering
	\includegraphics[width=\linewidth]{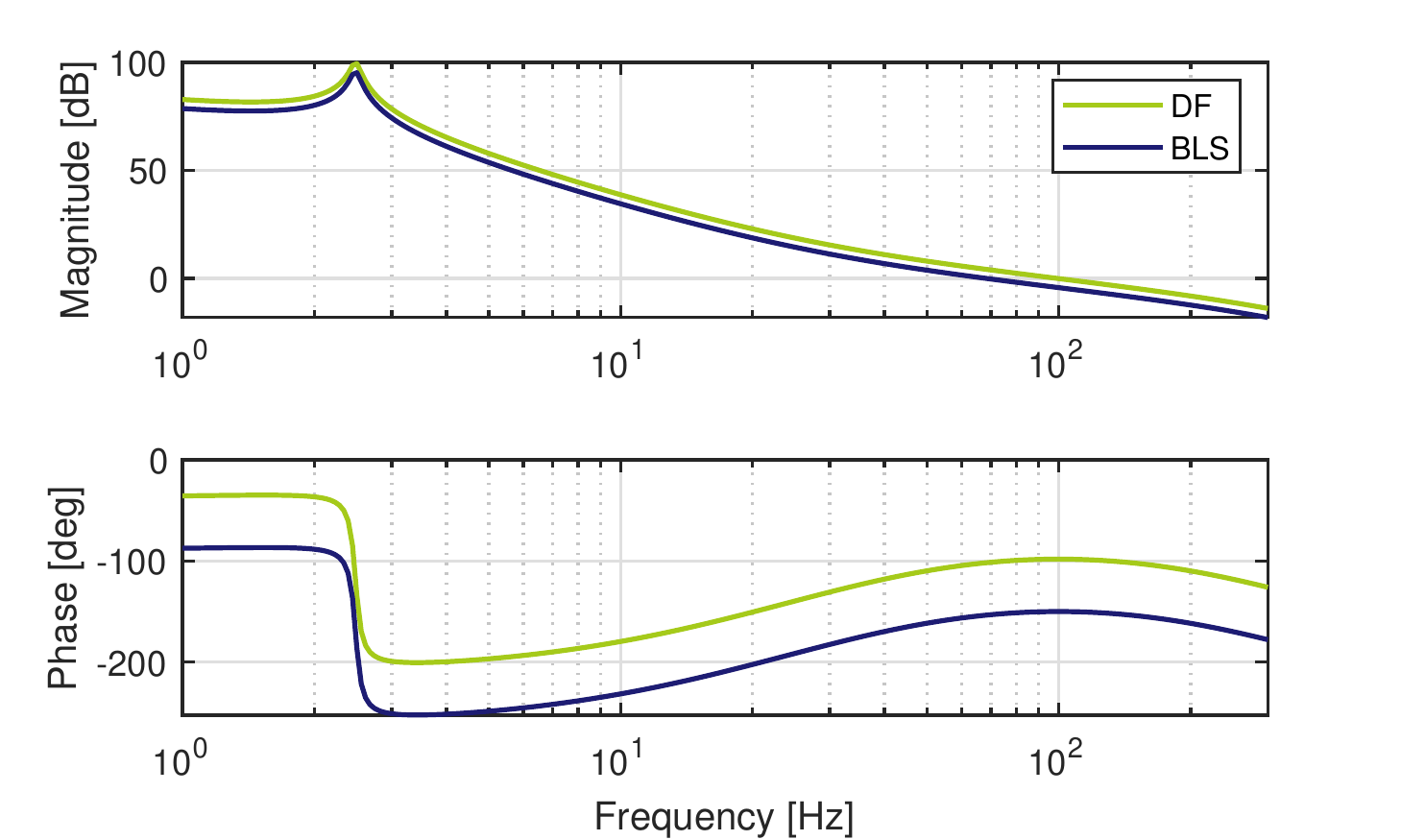}
	\caption{Bode plot of \eqref{eq:setup} controlled by a CI and a linear controller $C_{CI}$. The BLS and first harmonic values as computed using DF analysis are given.}
	\label{fig:RCIb} 
\end{figure}
\begin{example}[CI modelling]
	\par Potential errors caused by \assref{ass:xtr} are illustrated by means of a RCS with CI controller. Poor prediction performance is expected, as a CI often yields responses with many additional resets. Consider \eqref{eq:setup} with a fully resetting CI ($\gamma\;{=}\;0$) in series with the following PD\textsuperscript{2} controller:
	\begin{align}
	C_{CI}(s)\;=\;k_p\,\left(\frac{s\:{+}\:\omega_c\,{/}\,\beta}{s\:{+}\:\omega_c\,\beta}\right)^2,\quad \beta\;{=}\;3.73 \nonumber
	\end{align}
	where $k_p$ is adjusted to ensure $\omega_c\;{=}\;100$ Hz. This system has $PM_{BLS}\;{=}\;30^\circ$, with $\phi_{RC}\;{=}\;51.9^\circ$ added through CI \cite{Clegg1958}. The Bode plot is given in \figref{fig:RCIb}. As full reset is used, consecutive resets have negligible effects.
	\par The ISE and $L_\infty$ metrics are given for all three description methods by \figref{fig:RCIe}. ISE values in particular are found to be excessive, exceeding $100\,\%$ for many frequencies, for all methods. This is explained by the time response for a $20$ Hz reference, provided by \figref{fig:RCIt}. Without time regularization the simulated response is seen to have numerous resets per period, even affecting the weight of the modelled one.
	\par For full time regularization the simulated response visually coincides with the $\delta$-CL prediction. The marginal errors left must be caused by \assref{ass:2}. The CL-DF erroneously predicts its reset time. This prediction is consistent with its $\vec{q}_{DF,1}\;{=}\;0$ assumed reset time (see \tabref{tab:assumptions}), but deviates considerably from the simulated signal.
	\begin{figure}[t]
		\centering
		\includegraphics[width=\linewidth]{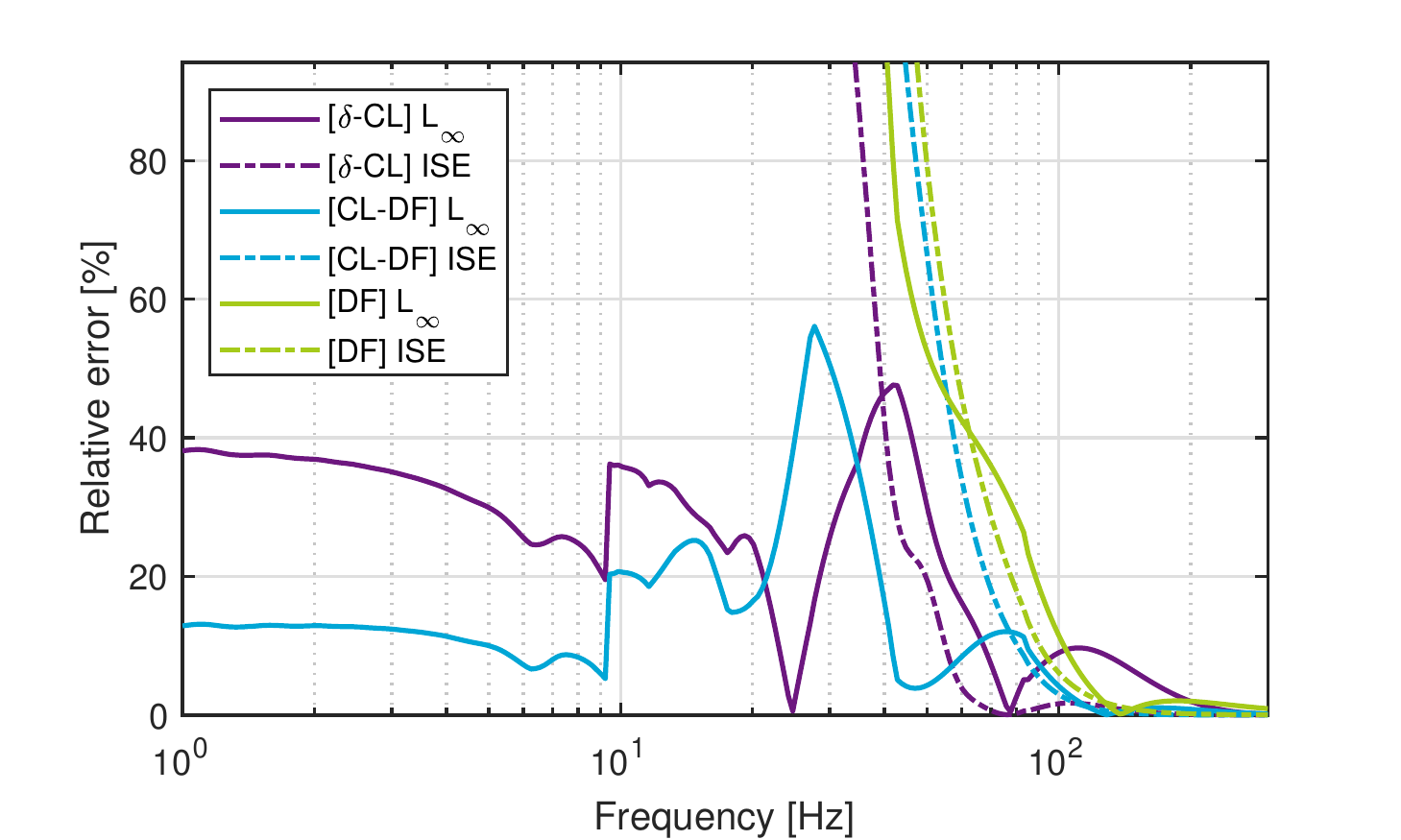}
		\caption{ISE and $L_\infty$ performance indicators for \eqref{eq:setup} controlled using a CI and $C_{CI}$. Results are shown for the three analytical describing methods.}
		\label{fig:RCIe} 
		\vspace{-3mm}
	\end{figure}
	\begin{figure}[t]
		\centering
		\includegraphics[width=\linewidth]{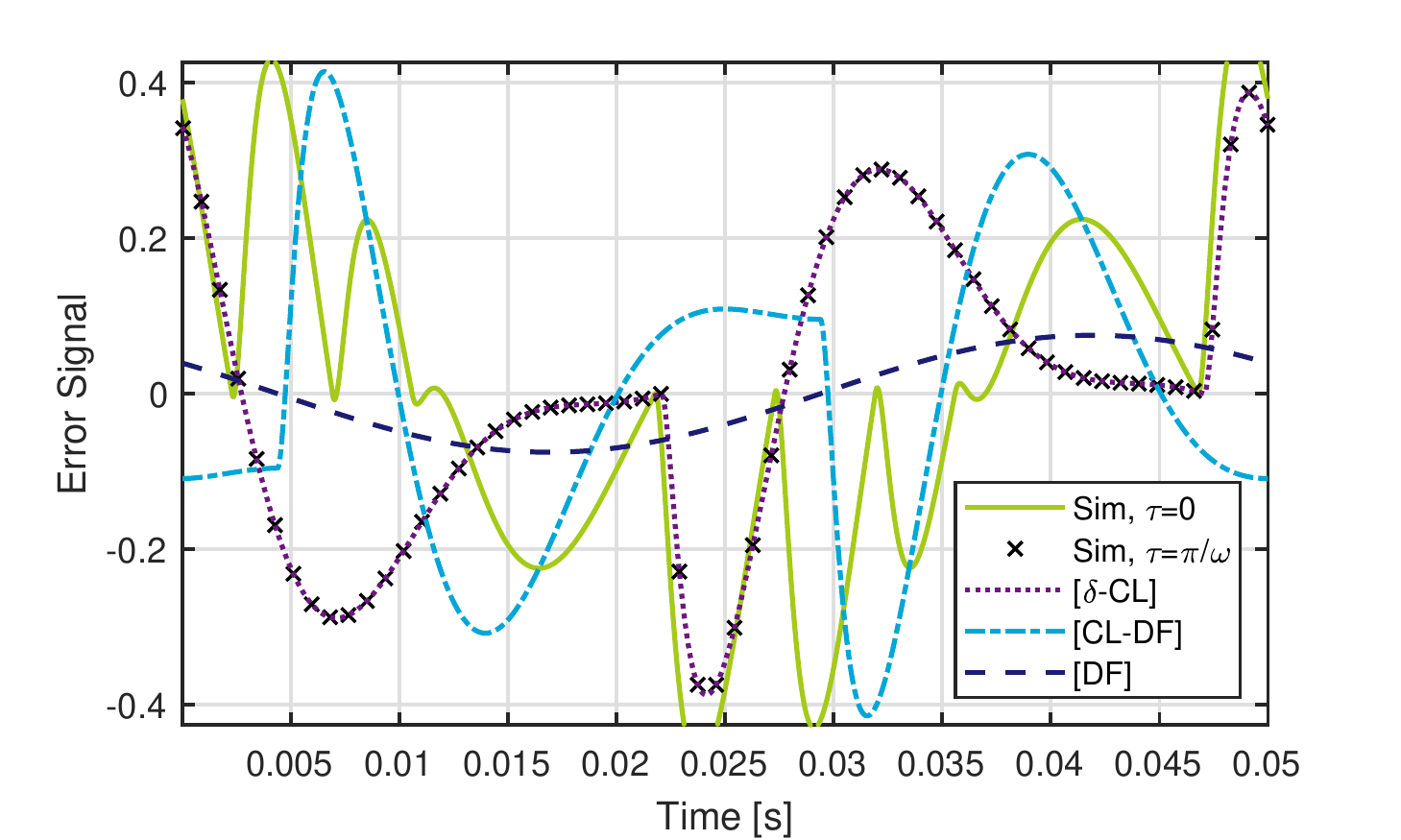}
		\caption{Error time responses for \eqref{eq:setup} controlled using a CI and a linear controller $C_{CI}$ with a $20$ Hz unit magnitude reference. Simulated responses are given with either no ($\tau\;{=}\;0$) or full ($\tau\;{=}\;\pi\,{/}\,\omega$) time regularization, in addition to modelled responses using the three analytical RC descriptions.}
		\label{fig:RCIt} 
		\vspace{-3mm}
	\end{figure}
\end{example}
\subsection{Effects of \assref{ass:2}}
\par The effects of \assref{ass:2} can be isolated by applying full time regularization. $\Phi\:{\ll}\:180^\circ$ is assumed. $\Phi$ is expected to be small if the combined effects of all prior resets have dampened out at a reset instant. It follows that:
\begin{itemize}
	\item More errors are expected at high frequencies, since there is less time between resets, thus less time for reset-induced impulses to dampen out.
	\item Higher errors are expected for a lower $PM_{BLS}$, as a lower $PM$ generally increases settling times.
\end{itemize}
\par
\begin{figure}[t]
	\centering
	\includegraphics[width=\linewidth]{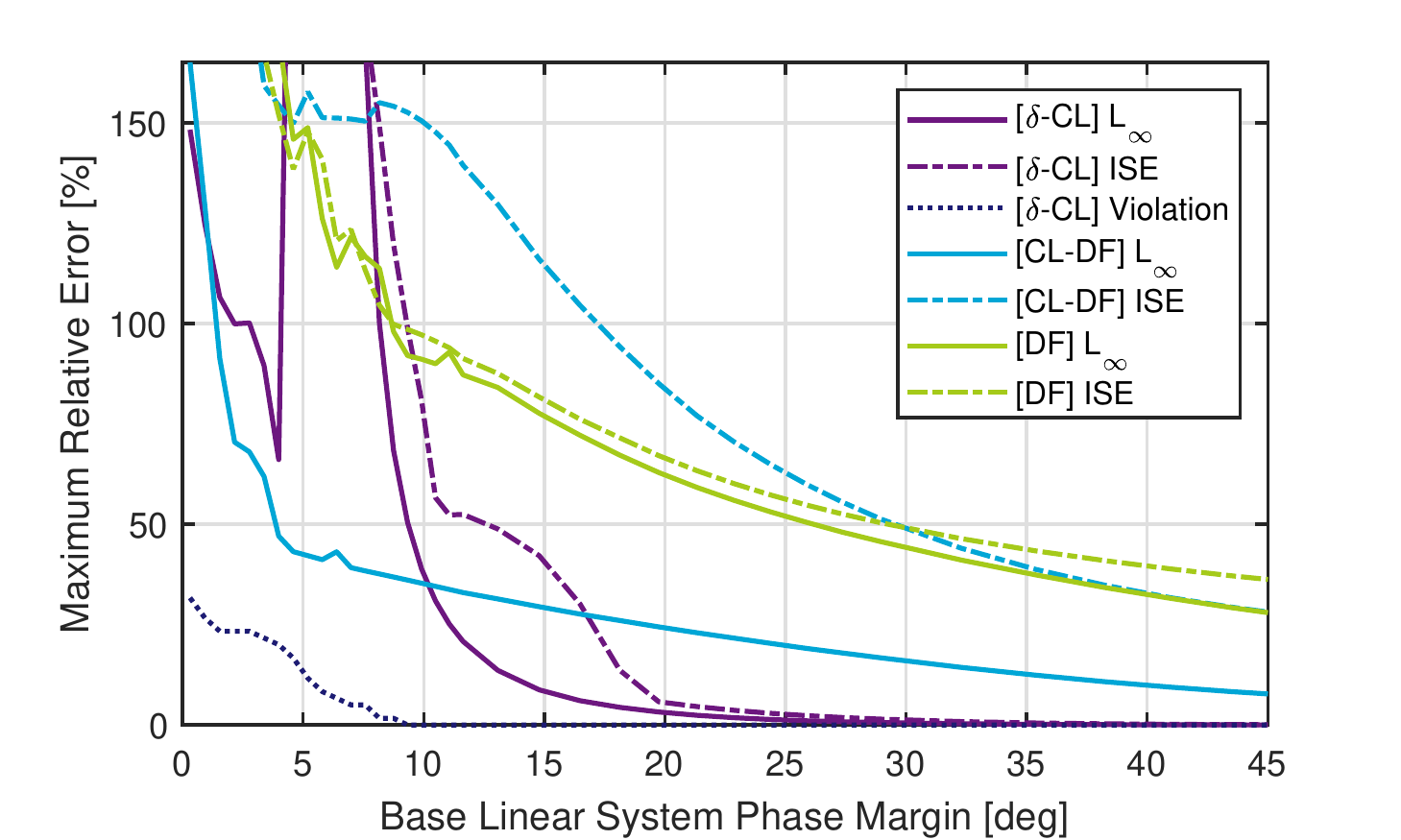}
	\caption{Performance metrics plotted against $PM_{BLS}$, where the worst performing frequency between $10$ and $100$ Hz is used, per value of $PM_{BLS}$. The range of $PM_{BLS}$ is obtained by sweeping $\beta$ from $4.5$ to $1.37$, while otherwise using $\mathcal{R}_1^\star$ on \eqref{eq:setup}. Full time regularization is used. The percentage of frequencies between $10$ and $100$ Hz violating \assref{ass:3} is given.}
	\label{fig:pm} 
	\vspace{-3mm}
\end{figure}

 \figref{fig:pm} gives the worst-case performance metrics when full time regularization is used, as a function of $PM_{BLS}$. \assref{ass:2} is found to be violated at very low values of $PM_{BLS}$. $\delta$-CL will thus give erroneous results there. For reasonable $PM_{BLS}$, \figref{fig:pm} shows that prediction errors due to \assref{ass:2} shrink when $PM_{BLS}$ increases. At most $PM_{BLS}$ points $\delta$-CL greatly outperforms the DF and CL-DF descriptions, both in ISE and $L_\infty$ terms. 
\vspace{0.5mm}
\par
\begin{figure}[b]
	\vspace{-4mm}
	\centering
	\includegraphics[width=\linewidth]{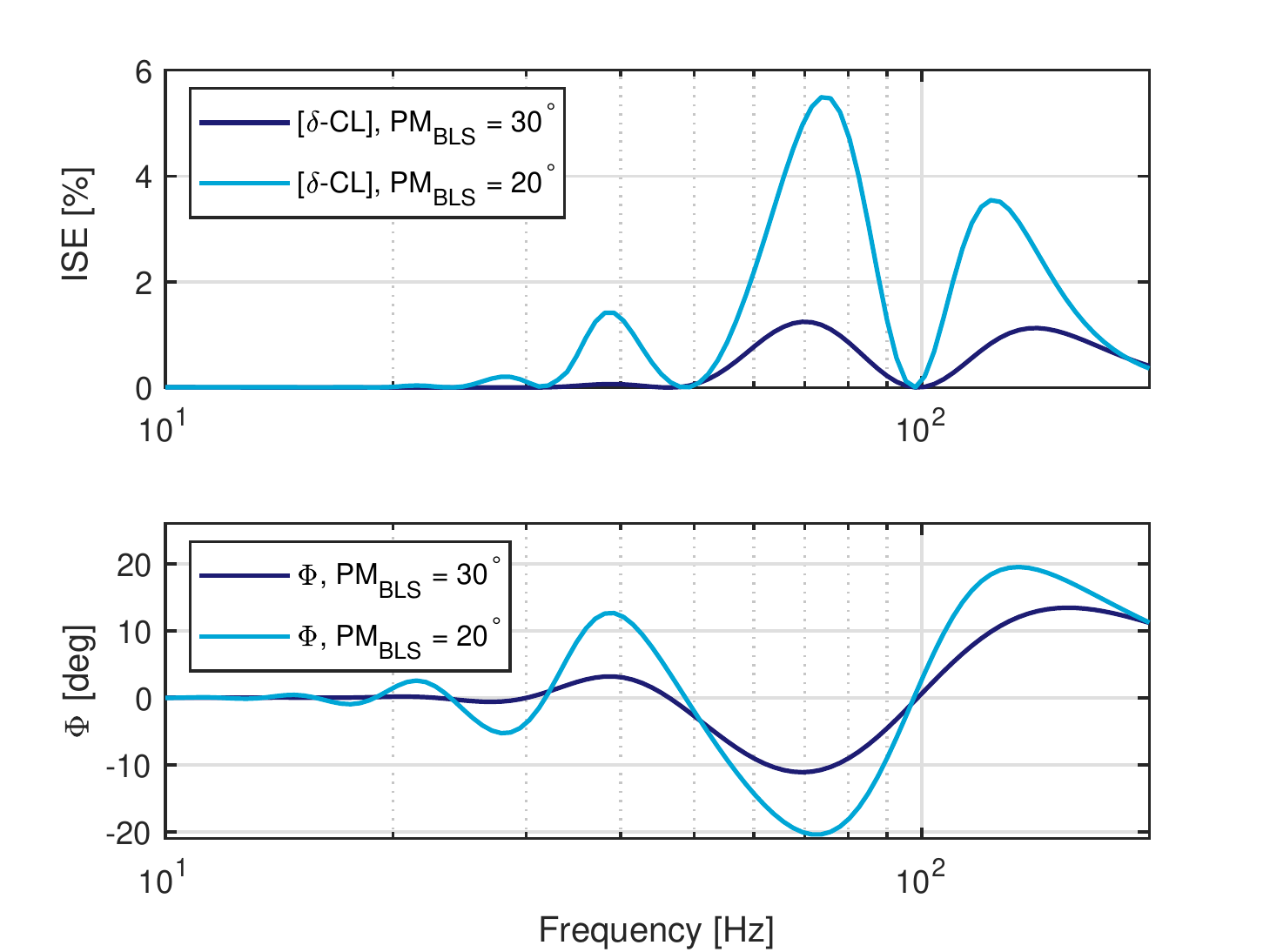}
	\caption{ISE and $\Phi$ values, for $\mathcal{R}^\star_1$ and $\mathcal{R}^\star_2$ acting on \eqref{eq:setup}, visualizing the relations between $\Phi$, $PM_{BLS}$ and ISE. Full time regularization is used.}
	\label{fig:pm2} 
\end{figure}

 \figref{fig:pm2} provides ISE and $\Phi$ results for two controllers with different $PM_{BLS}$ as function of frequency, using full time regularization. As such, all modelling errors are caused by \assref{ass:2}, which means that the solutions should be exact if $\Phi\;{=}\;0$. This holds, as can be seen in \figref{fig:pm2}.
\subsection{Effects of \assref{ass:5}}
\par
\begin{figure}[t]
	\centering
	\includegraphics[width=\linewidth]{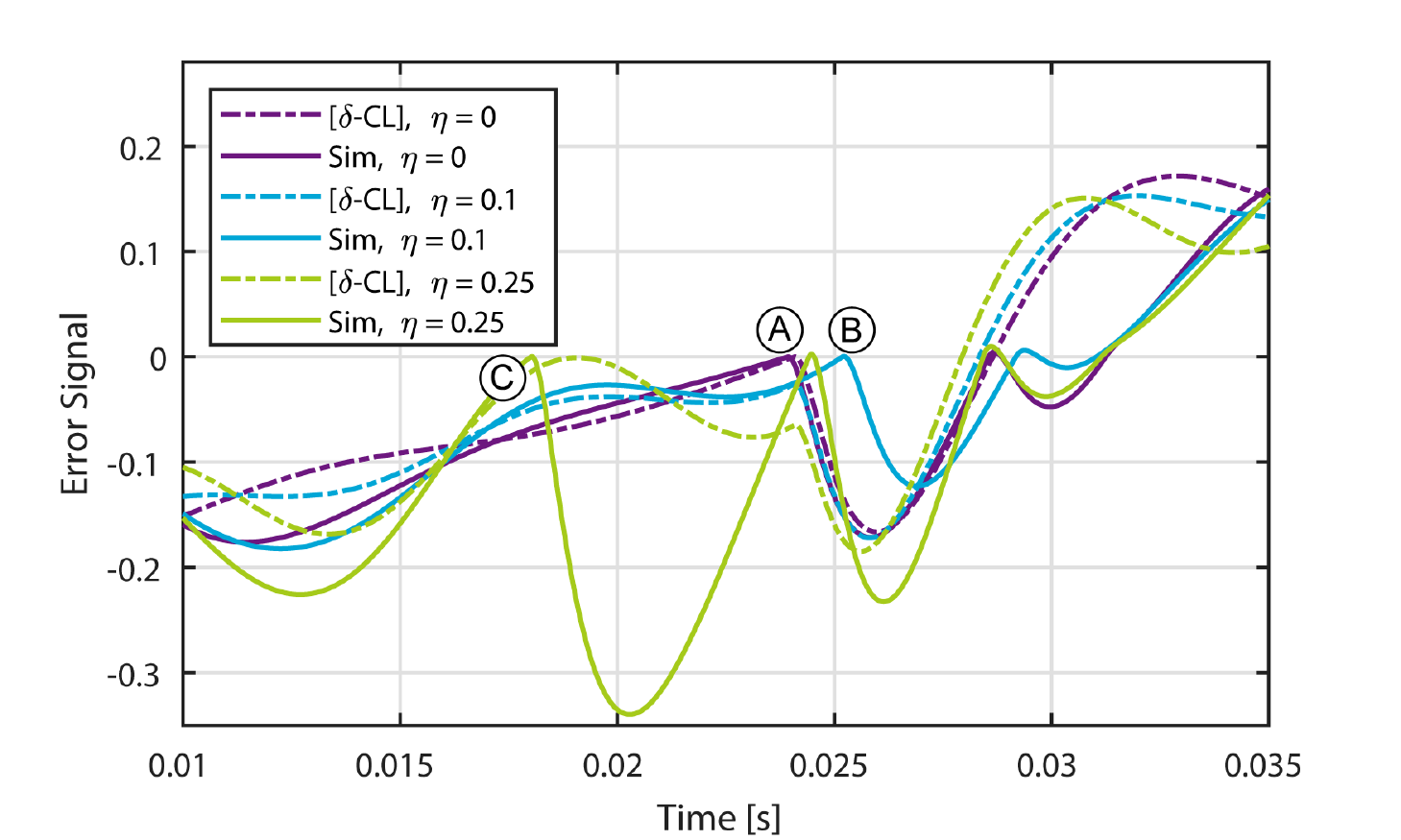}
	\caption{Time domain simulations and $\delta$-CL predictions for a $20$ Hz reference and a $100$ Hz disturbance with magnitudes $\eta$ relative to the reference, using controller $\mathcal{R}_0^\star$ on \eqref{eq:setup}. No time regularization is used. A half period response is shown. The disturbance phase is chosen such that its peaks coincide with zero-crossings of $\vec{q}(t)$ when there would not be a disturbance.}
	\label{fig:multsine} 
\end{figure}

 This assumption extends the generality of $\delta$-CL by permitting multi-sine references and disturbances. One of the worst cases for this assumption is if some other input has its peaks coinciding with the zero crossings of the base sinusoid, as this causes it to have a considerable effect on the reset times. Consider disturbance $\vec{d}$, introduced between $C(s)$ and $P(s)$, having a phase computed to meet this worst case scenario and magnitude parametrized by $\eta\;{=}\;\vert \vec{d} \vert\,{/}\,\vert \vec{r}_I \vert$.  \figref{fig:multsine} gives sample predictions and simulations for various $\eta$.
\par The system without disturbance ($\eta\;{=}\;0$) has a reset at (A), which is the one modelled by $\delta$-CL. \assref{ass:5} considers the corresponding reference to be the only one causing resets. Predictions with disturbance therefore also model their resets at (A). However, the reset for $\eta\;{=}\;0.1$ occurs at (B), causing a slight error in predicted peak position. The case for $\eta\;{=}\;0.25$ is far worse, as it generates an additional reset at (C), which is not captured at all yet inflicts the largest error peak.
\par The effects of \assref{ass:5} depend on numerous parameters, including plant model, controller and input type. Its validity should therefore be evaluated per individual system.
\vspace{0.5mm}
\subsection{Method validity}
\par The various error sources are found to be well-defined, as they can be clearly linked to Assumptions \shortref{ass:xtr} or \shortref{ass:2}. In case both assumptions hold all modelling errors equal zero. As such the solution without assumptions, \eqref{eq:Ehosidf}, is exact, which is as expected based on the mathematical derivation.
\par $\delta$-CL is an approximation, relying on Assumptions \shortref{ass:xtr} and \shortref{ass:2}. The former rarely holds in practice, yet does not necessarily inflict large errors, as additional resets are generally of lower magnitude than the modelled ones. Exceptions exist, as seen in \figref{fig:RCIt} for $\tau\;{=}\;0$. Increasing $\tau$ diminishes these errors.
\par \assref{ass:2} holds if $PM_{BLS}$ is sufficiently large, in which case it inflicts small errors relative to \assref{ass:xtr}. Based on \figref{fig:pm}, a $PM_{BLS}\:{\gtrapprox}\:20^\circ$ is advised for using $\delta$-CL. These limits are system dependent.
\section{Simulation Results}
\label{sec:res}
\par The performance of $\delta$-CL is further examined and compared to that of CL-DF and DF using various CgLp tunings, provided by \tabref{tab:CgLptuning}. Optimal time regularization is used. \tabref{tab:results} tabulates the peak and log-space average ISE and $L_\infty$ metrics. From \tabref{tab:results} it can be concluded that, in terms of ISE, $\delta$-CL consistently outperforms CL-DF and DF. In terms of $L_\infty$, differences between $\delta$-CL and CL-DF are less pronounced, with either method yielding similar results, though both performing significantly better than DF. As expected, prediction errors increase when $PM_{BLS}$ decreases. From comparing mean and median values for optimal and full time regularization it is concluded that the main error source for $\delta$-CL must be unmodelled resets, as prediction errors reduce significantly if full time regularization is applied. The same does not hold for CL-DF nor DF, which retain similar ISE and $L_\infty$ values when removing additional resets.

\begin{table}[t]
	\centering
	\caption{CgLp and PID controller details with $\phi_{RC}$ indicating the phase lead provided through reset at bandwidth as computed using DF analysis. For all controllers $\omega_i\;{=}\;10$ Hz, $\omega_c\;{=}\;100$ Hz and $\omega_f\;{=}\;500$ Hz. Gain $k_p$ is adjusted to achieve bandwidth $\omega_c$.}
	\label{tab:CgLptuning}
	\begin{tabular}{lcccccc}
		\hline
		& & & & & & \\  [-0.8em]
		& $PM_{BLS}$ & $\phi_{RC}$ & $\gamma$ & $\omega_r$ [Hz] & $\alpha$ & $\beta$ \\ \hline
		& & & & & & \\  [-0.8em]
		${\mathcal{R}_0}$ & $20^\circ$ & $40^\circ$ & $0$        	& $34.41$ & $1.24$ & $2.17$  \\
		${\mathcal{R}_1}$ & $30^\circ$ & $30^\circ$ & ${-}\:0.2$ 	& $83.46$ & $1.18$ & $2.87$  \\
		${\mathcal{R}_2}$ & $30^\circ$ & $30^\circ$ & $0$ 			& $62.88$ & $1.15$ & $2.78$  \\
		${\mathcal{R}_3}$ & $30^\circ$ & $30^\circ$ & $0.2$ 		& $37.55$ & $1.12$ & $2.68$  \\
		${\mathcal{R}_4}$ & $40^\circ$ & $20^\circ$ & $0$ 			& $98.38$ & $1.07$ & $3.59$  \\
		${\mathcal{R}_5}$ & $40^\circ$ & $30^\circ$ & $0$ 			& $62.88$ & $1.15$ & $3.79$  \\
		${\mathcal{R}_6}$ & $50^\circ$ & $30^\circ$ & $0$ 			& $62.88$ & $1.15$ & $5.79$  \\
		${\mathcal{R}_7}$ & $50^\circ$ & $40^\circ$ & $0$ 			& $34.41$ & $1.24$ & $5.81$  \\ 
		%${\mathcal{R}_8}$ & $40^\circ$ & $50^\circ$ & $-0.2$		& $33.35$ & $1.44$ & $4.07$  \\
		%${\mathcal{R}_9}$ & $30^\circ$ & $60^\circ$ & $-0.4$		& $33.31$ & $1.80$ & $3.16$  \\ 
		\hline	
	\end{tabular}
\end{table}

\begin{figure}[b]
	\vspace{-3mm}
	\centering
	\includegraphics[width=\linewidth]{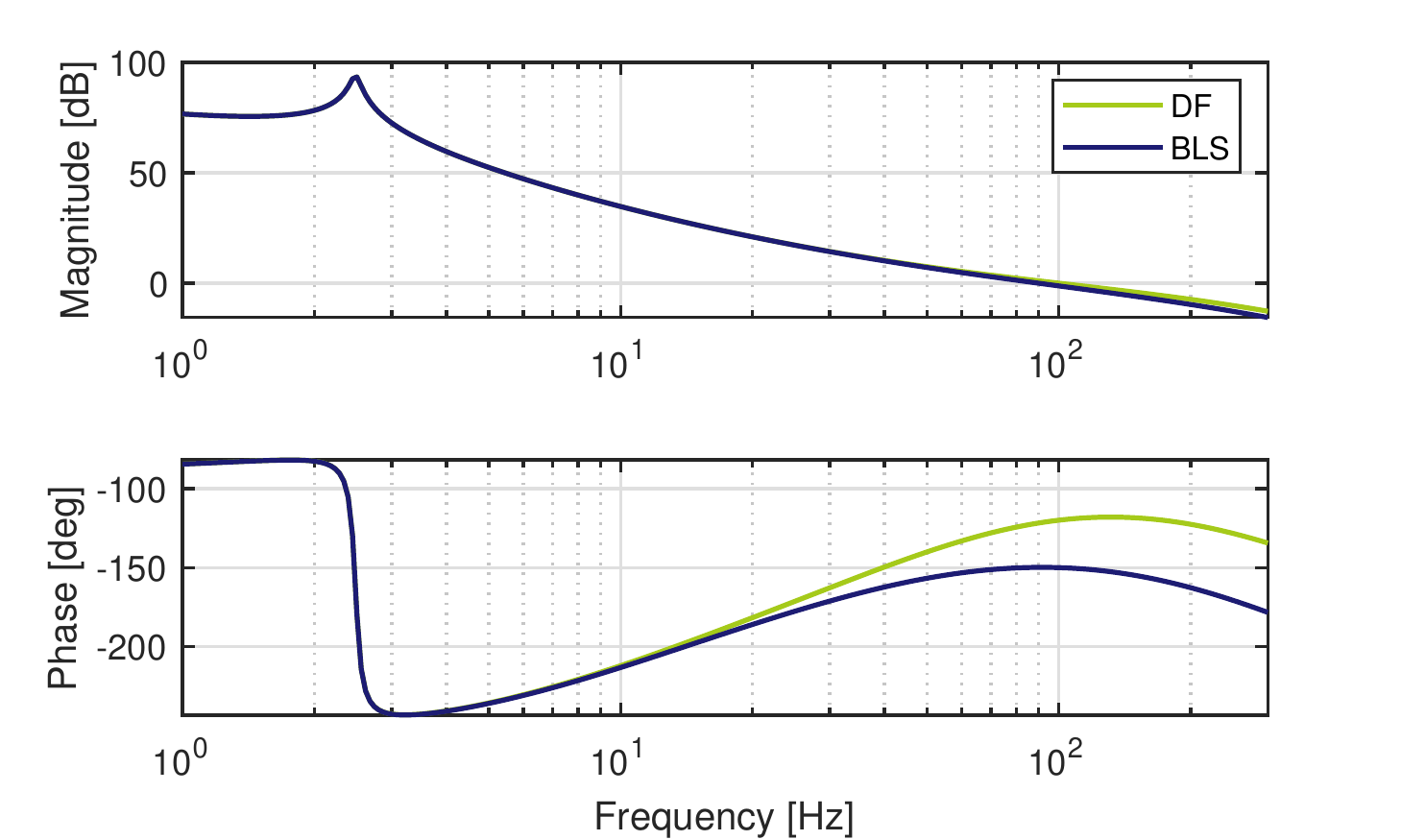}
	\caption{Bode Plot for plant \eqref{eq:setup} with controller $\mathcal{R}_2$. The responses of the BLS and nonlinear RC modelled through DF analysis are shown.}
	\label{fig:r2bode} 
\end{figure}
\subsection{Examining results for $\mathcal{R}_2$}
	\par Results for $\mathcal{R}_2$ are analysed in greater detail. This controller is selected because it roughly represents the median of all controllers in terms of performance. \figref{fig:r2bode} provides the Bode plot for this setup, showing that the $PM_{BLS}$ is $30^\circ$ and that reset adds $\phi_{RC}\;{=}\;30^\circ$, as predicted through DF analysis.
	\par
\begin{figure}[t]
		\vspace{-4mm}
		\centering
		\includegraphics[width=\linewidth]{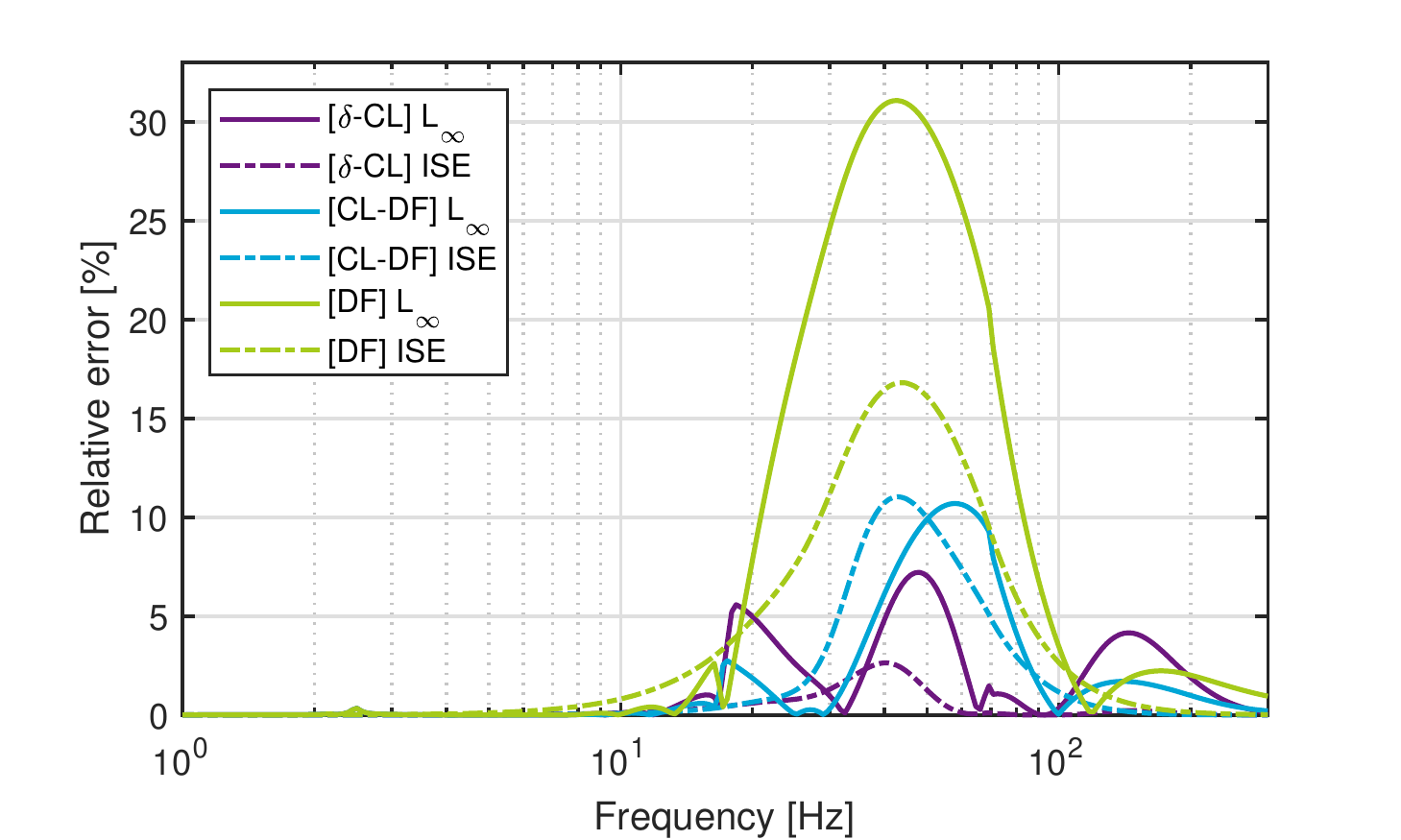}
		\caption{$L_\infty$ and ISE performance indicators for all three prediction methods over a range of frequencies, using plant \eqref{eq:setup} with controller $\mathcal{R}_2$. Optimal time regularization ($\tau\;=\;2\pi\,{/}\,(10\,\omega_c)$) is used.}
		\label{fig:r2error} 
	\end{figure}

 \figref{fig:r2error} shows the performance metrics for all three prediction methods. In terms of ISE, $\delta$-CL outperforms CL-DF and DF. More subtle differences between $\delta$-CL and CL-DF are found for the $L_\infty$ metric, with either method improving upon the other in some frequency range.	
	\begin{figure}[t]
		\vspace{-4mm}
		\centering
		\includegraphics[width=\linewidth]{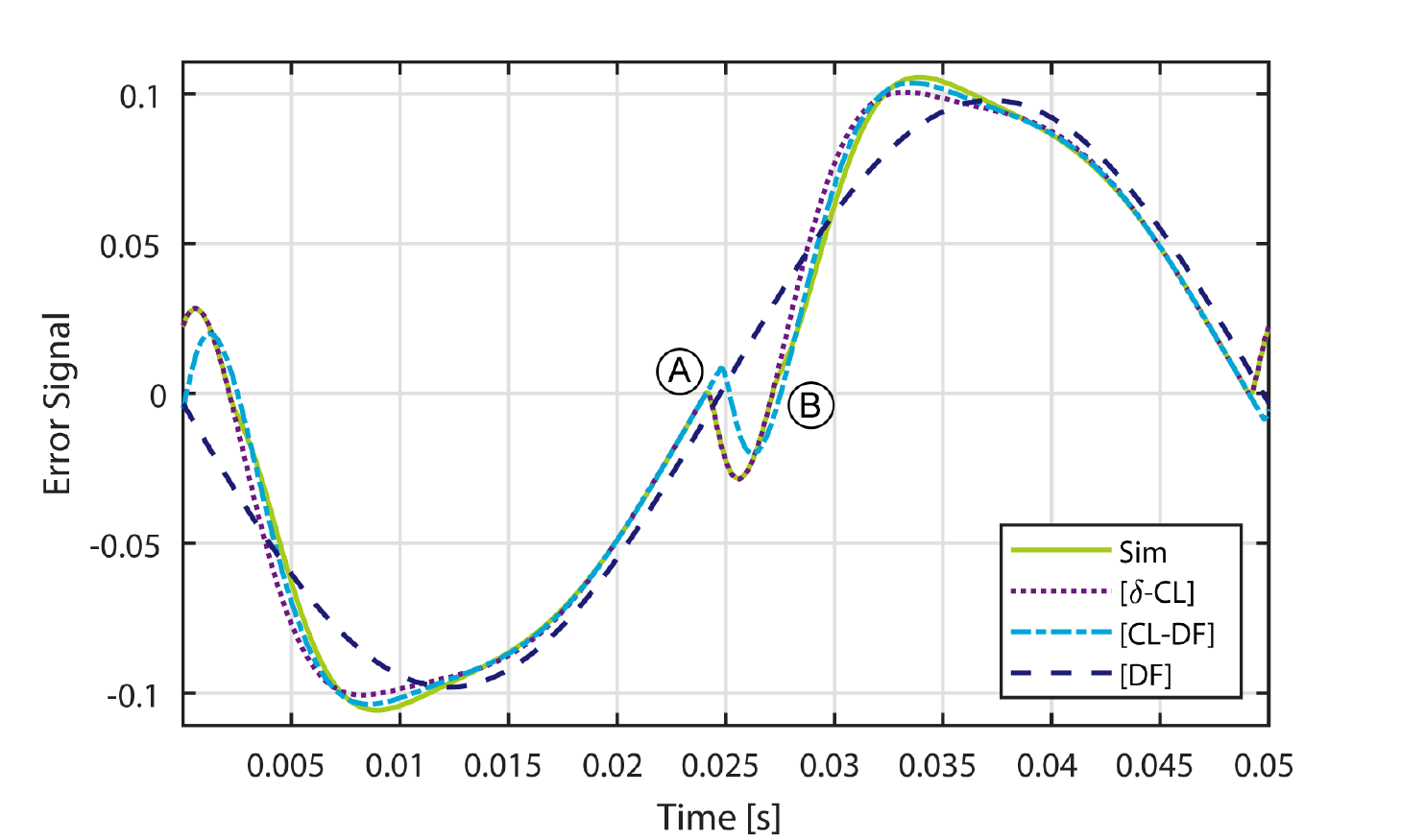}
		\caption{Simulated and predicted error time responses for plant \eqref{eq:setup} with controller $\mathcal{R}_2$, using a $20$ Hz sinusoidal $\vec{r}_I(t)$ with unit magnitude. The reset instants for a half period, (A) and (B), are indicated.}
		\label{fig:r220hz} 
		\vspace{-4mm}
	\end{figure}
\begin{table*}[t]
	\centering
	\caption{Analytic RCS description method performance using plant \eqref{eq:setup} with the controllers in \tabref{tab:CgLptuning}. Optimal time regularization is used. The relative peak and log-space average ISE and $L_\infty$ results are given, evaluated between $1$ Hz and bandwidth. Mean and median values are given for both optimal and full time regularization.}
	\label{tab:results}
	\begin{adjustbox}{angle=90}
	\begin{tabular}{lccccccccccccccc}
		& \multicolumn{3}{c}{$\text{max}_{\omega}\,$ISE} && \multicolumn{3}{c}{$\text{max}_{\omega}\,L_\infty$} && \multicolumn{3}{c}{$\text{avg}_{\omega}\,$ISE} && \multicolumn{3}{c}{$\text{avg}_{\omega}\,L_\infty$} \\ \hline
		& & & & & & & & & & & & & & & \\  [-0.8em]
		& $\delta$-CL & CL-DF      & DF       && $\delta$-CL        & CL-DF      & DF      && $\delta$-CL & CL-DF      & DF       && $\delta$-CL        & CL-DF      & DF    \\ \hline
		& & & & & & & & & & & & & & & \\  [-0.8em]
		& \multicolumn{15}{c}{\textit{Results for optimal time regularization:}}\\
		\hline
		& & & & & & & & & & & & & & & \\  [-0.8em]
		${\mathcal{R}_0}$ & $69.0\:\%$  & $94.0\:\%$ & $114\:\%$ && $33.1\:\%$ & $19.0\:\%$  & $94.1\:\%$ && $12.5\:\%$  & $16.1\:\%$  & $23.6\:\%$ && $5.30\:\%$  & $4.35\:\%$  & $22.6\:\%$ \\ 
		${\mathcal{R}_1}$ & $1.33\:\%$  & $7.82\:\%$ & $12.9\:\%$ && $8.71\:\%$ & $15.2\:\%$   & $29.1\:\%$ && $0.113\:\%$  & $1.37\:\%$  & $3.03\:\%$ && $0.719\:\%$ & $1.90\:\%$  & $6.61\:\%$ \\
		${\mathcal{R}_2}$ & $2.63\:\%$  & $11.0\:\%$ & $16.8\:\%$ && $7.22\:\%$ & $10.7\:\%$  & $31.1\:\%$ && $0.383\:\%$  & $1.93\:\%$  & $4.13\:\%$ && $1.20\:\%$  & $1.78\:\%$  & $7.62\:\%$ \\
		${\mathcal{R}_3}$ & $11.1\:\%$  & $25.5\:\%$ & $31.4\:\%$ && $15.7\:\%$ & $9.83\:\%$  & $40.9\:\%$ && $2.56\:\%$  & $4.72\:\%$  & $8.03\:\%$ && $2.91\:\%$  & $1.53\:\%$  & $10.6\:\%$ \\
		${\mathcal{R}_4}$ & $0.00720\:\%$ & $1.21\:\%$ & $3.26\:\%$ && $0.262\:\%$& $4.07\:\%$   & $14.0\:\%$ && $0.00120\:\%$ & $0.215\:\%$  & $0.791\:\%$ && $0.0412\:\%$& $0.551\:\%$ & $2.87\:\%$ \\
		${\mathcal{R}_5}$ & $1.63\:\%$  & $7.77\:\%$ & $12.6\:\%$ && $5.94\:\%$  & $5.55\:\%$  & $22.6\:\%$ && $0.287\:\%$  & $1.59\:\%$  & $3.34\:\%$ && $0.963\:\%$ & $1.13\:\%$  & $6.02\:\%$ \\
		${\mathcal{R}_6}$ & $1.17\:\%$  & $6.45\:\%$ & $10.0\:\%$ && $4.89\:\%$ & $4.25\:\%$  & $17.0\:\%$ && $0.238\:\%$  & $1.46\:\%$  & $2.83\:\%$ && $0.814\:\%$ & $0.610\:\%$ & $4.96\:\%$ \\
		${\mathcal{R}_7}$ & $12.9\:\%$  & $38.0\:\%$ & $40.0\:\%$ && $13.7\:\%$ & $5.62\:\%$  & $30.2\:\%$ && $3.26\:\%$  & $10.5\:\%$  & $12.3\:\%$ && $3.32\:\%$  & $1.37\:\%$  & $10.2\:\%$ \\
		%${\mathcal{R}_8}$ & $41.6\:\%$  & $96.9\:\%$ & $118\:\%$ && $25.4\:\%$ & $39.5\:\%$  & $107\:\%$ && $9.43\:\%$  & $23.9\:\%$  & $31.3\:\%$ && $4.85\:\%$  & $8.17\:\%$  & $26.2\:\%$ \\
		%${\mathcal{R}_9}$ & $224\:\%$   & $234\:\%$  & $554\:\%$  && $98.7\:\%$ & $88.4\:\%$   & $302\:\%$ && $42.6\:\%$  & $56.9\:\%$  & $107\:\%$ && $17.9\:\%$ & $21.3\:\%$   & $68.0\:\%$ \\ 
		\hline
		& & & & & & & & & & & & & & & \\  [-0.8em]
		%\textit{mean:} 		    	 & $40.0\:\%$  & $42.4\:\%$ & $66.7\:\%$ && $21.3\:\%$  & $20.2\:\%$ & $68.8\:\%$ && $10.9\:\%$  & $14.0\:\%$  & $22.5\:\%$ && $3.81\:\%$  & $4.27\:\%$ & $16.6\:\%$ \\  
		%\textit{median:}	    	 & $21.1\:\%$  & $28.4\:\%$ & $45.6\:\%$ && $11.2\:\%$ & $10.3\:\%$ & $30.6\%$ && $5.41\:\%$  & $8.16\:\%$  & $15.4\:\%$ && $2.06\:\%$  & $1.65\:\%$ & $8.90\:\%$ \\ \hline
		\textit{mean:} 		    	 & $12.5\:\%$  & $24.0\:\%$ & $30.1\:\%$ && $11.2\:\%$  & $9.28\:\%$ & $34.9\:\%$ && $2.38\:\%$  & $4.74\:\%$  & $7.26\:\%$ && $1.91\:\%$  & $1.65\:\%$ & $8.94\:\%$ \\  
		\textit{median:}	    	 & $2.13\:\%$  & $9.43\:\%$ & $14.9\:\%$ && $7.96\:\%$ & $7.72\:\%$ & $29.7\%$ && $0.335\:\%$  & $1.76\:\%$  & $3.73\:\%$ && $1.08\:\%$  & $1.45\:\%$ & $7.12\:\%$ \\ 
		\hline
		& & & & & & & & & & & & & & & \\  [-0.8em]
		& \multicolumn{15}{c}{\textit{Results for full time regularization:}}\\
		\hline
		& & & & & & & & & & & & & & & \\  [-0.8em]
		\textit{mean:} 		    	 & $2.65\:\%$  & $48.7\:\%$ & $42.0\:\%$ && $2.62\:\%$  & $20.4\:\%$ & $47.8\:\%$ && $0.181\:\%$  & $7.28\:\%$  & $8.63\:\%$ && $0.220\:\%$  & $3.12\:\%$ & $10.6\:\%$ \\  
		\textit{median:}	    	 & $0.0358\:\%$  & $13.6\:\%$ & $17.0\:\%$ && $0.620\:\%$ & $14.6\:\%$ & $36.8\%$ && $0.00195\:\%$  & $2.23\:\%$  & $3.84\:\%$ && $0.0438\:\%$  & $2.34\:\%$ & $7.30\%$ \\ 
		\hline
	\end{tabular}
	\end{adjustbox}
\vspace{-1mm}
\end{table*}
\par A time domain response is used for further analysis, given by \figref{fig:r220hz} for a $\omega\;{=}\;20$ Hz reference. At this frequency \figref{fig:r2error} indicates that, when considering $L_\infty$, CL-DF outperforms $\delta$-CL. The time domain responses show that $\delta$-CL captures the effects of reset instant (A) accurately. Reset (B), considerably smaller in magnitude, is not modelled, causing the found ISE and $L_\infty$ errors. Considering CL-DF, \figref{fig:r220hz} shows that it models the response to (A) with some offset. While this causes ISE errors, this erroneous placement works to its advantage here, as the incorrect positioning of (A) compensates for not modelling (B) in evaluating the peak error, $L_\infty$. As such, $L_\infty$ performance of CL-DF is, in this case, better than that of $\delta$-CL.
\par The erroneous impulse position prediction of CL-DF does not generally works to its advantage. \figref{fig:RCIt} gives an example where large ISE errors are inflicted by the CL-DF reset placement assumption. Evaluating \eqref{eq:E} shows that the error signal equals the BLS with added impulse responses. Thus, zero crossings occur near those of the BLS, provided that the impulse responses are sufficiently convergent. This supports that the reset placement assumption used by $\delta$-CL is often more accurate than that of CL-DF.

\section{Conclusion}
\label{sec:con}
% Paragraph on advantage of RC and limitations of describing methods
\par Reset control can overcome fundamental limitations of linear control, whilst permitting design using the industry-preferred loop-shaping methodology. Even though accurate tuning through loop-shaping requires a thorough understanding of closed-loop behaviour, no frequency-domain methods found in literature sufficiently describe the principles of how open-loop reset control design translates to closed-loop behaviour. Additionally, no methods linking the base-linear system design to the closed-loop RCS performance is found.
% Paragraph with paper contribution: impulse formulation (MIMO or different reset laws) that needs numerical solving, exact closed-loop description which needs to be solved numerically, analytical method with clear error sources.
\par A rarely mentioned approach, which models open-loop reset control as a linear system with state-dependent impulse train inputs, is taken and generalized. It is shown that any generic closed-loop reset system behaves as the base-linear system with added impulse responses. This describes, to the authors' knowledge, for the first time the underlying principles that link open-loop reset control design to its closed-loop performance, using frequency-domain terms as required for loop-shaping. This insight may be used to improve reset control design.
\par An analytical solution is obtained by inserting some well-defined assumptions. This description is critically examined using simulations and compared to existing methods. The novel description consistently provides a considerably more accurate time-domain prediction than the best performing analytical method found in literature, whilst having a similar performance in terms of predicted peak errors. It is shown how the various assumptions contribute to the prediction error, giving a good understanding of method limitations. High accuracy is attained if the base-linear system has sufficient phase margin and if unmodelled resets have a comparatively low magnitude, conditions met by various practical implementations.
% Conclusion sentence on application
\par Concluding, it is found that the presented description provides additional insight and improves predictions for reset control analysis, which can be used to improve reset element design. Further research should be conducted on how these results can be applied to practical reset control tuning.

% Bibliography
\bibliographystyle{plain}
\bibliography{References}

\end{document}